\documentclass{aa}  
\usepackage{graphicx}
\usepackage{txfonts}
\usepackage[colorlinks=true,citecolor=blue]{hyperref}
\usepackage{array}  
\usepackage{booktabs}  
\usepackage{multirow}  

\usepackage{placeins}           

\newcommand{\HFR}{HFR}
\newcommand{\Rossbyno}{${n=1}$~equatorial~Rossby}
\newcommand{\Rossby}{${n=0}$~equatorial~Rossby}

\usepackage{xcolor}
\usepackage{amssymb}
\usepackage{amsmath} 
\usepackage{mathrsfs} 
\usepackage{stmaryrd}
\usepackage{dsfont}
\usepackage[T1]{fontenc}
\usepackage{txfonts}
\usepackage{bm}
\usepackage{mathtools}
\usepackage{empheq}
\usepackage{comment}
\usepackage{natbib,twoopt}
\bibpunct{(}{)}{;}{a}{}{,}             
\makeatletter
  \newcommandtwoopt{\citeads}[3][][]{\href{http://adsabs.harvard.edu/abs/#3}%
    {\def\hyper@linkstart##1##2{}%
     \let\hyper@linkend\@empty\citealp[#1][#2]{#3}}}
  \newcommandtwoopt{\citepads}[3][][]{\href{http://adsabs.harvard.edu/abs/#3}%
    {\def\hyper@linkstart##1##2{}%
     \let\hyper@linkend\@empty\citep[#1][#2]{#3}}}
  \newcommandtwoopt{\citetads}[3][][]{\href{http://adsabs.harvard.edu/abs/#3}%
    {\def\hyper@linkstart##1##2{}%
     \let\hyper@linkend\@empty\citet[#1][#2]{#3}}}
  \newcommandtwoopt{\citeyearads}[3][][]%
    {\href{http://adsabs.harvard.edu/abs/#3}
    {\def\hyper@linkstart##1##2{}%
     \let\hyper@linkend\@empty\citeyear[#1][#2]{#3}}}
    \renewcommand*\aa@pageof{, page \thepage{} of \pageref*{LastPage}}
\makeatother

\begin{document}

   \title{Assessing the validity of the anelastic and Boussinesq approximations to model solar inertial modes}
\author{Suprabha Mukhopadhyay \inst{1}
\and Yuto Bekki \inst{1}
\and Xiaojue Zhu \inst{1}
\and Laurent Gizon \inst{1,2}} 
\institute{Max-Planck-Institut f{\"u}r Sonnensystemforschung,
              Justus-von-Liebig-Weg 3, 37077 G{\"o}ttingen, Germany
         \and
         Institut f{\"u}r Astrophysik und Geophysik, Georg-August-Universt{\"a}t G{\"o}ttingen,
         Friedrich-Hund-Platz 1, 37077 G{\"o}ttingen, Germany\\
         \email{\href{mailto:gizon@mps.mpg.de}{gizon@mps.mpg.de}},  \email{\href{mailto:zhux@mps.mpg.de}{zhux@mps.mpg.de}}}

    \authorrunning{Mukhopadhyay et al.}
   \date{Received / accepted }

\abstract
{Global-scale inertial modes of oscillations have been recently observed on the Sun. They might play an important dynamic and diagnostic role for the Sun. 
}
{This work aims to assess the validity of simplifying assumptions in the continuity equation, which have often been used in the linear models of solar inertial modes.
}
{{We compute the linear eigenmodes of the Sun's convection zone in the inertial frequency range using the Dedalus code. This single framework enables us to compare the sensitivity of the modes to different model setups, such as the compressible setup and the  Boussinesq and anelastic approximations. We consider both the cases of uniform rotation and solar differential rotation (as given by helioseismology).}}
{{We find that the compressible and anelastic models have almost identical eigenmodes under uniform and solar differential rotation. On the other hand, the absence of density stratification in the Boussinesq model results in significantly different eigenmodes under this formulation. The differences are most prominent for the non-toroidal modes with significant radial motions mainly due to the absence of the compressional $\beta$~effect.}}
{The anelastic approximation simplifies the calculations and reduces the numerical cost without affecting the solar inertial modes. The Boussinesq or incompressible approximations cannot be used to model the solar inertial modes accurately. {Given the strong influence of differential rotation on the eigenmodes, an acceptable setup is to use the anelastic approximation together with the solar differential rotation.}
} 


   \maketitle
%

    \begin{table*}
\small
    \caption{Classes of solar inertial modes studied in this paper.}
    \centering
    \begin{tabular}{c c c c c c}
    
    \hline\hline
      mode names used here    & other names & observed     & propagation direction&   north-south symmetry & linear models \\ 
      && on the Sun? &in Carrington frame &of radial vorticity&\\
        \hline
         \Rossby   & ... & yes (1--4)  & retrograde & symmetric & (1), (7--11) \\
         high-latitude inertial   & ... & yes (5)   & retrograde & both symmetries  & (5), (9) \\ 
         \HFR    & mixed modes  & yes (6)  & retrograde & anti-symmetric  & (11--13) \\ \hline
         \Rossbyno & mixed modes & maybe (4) & retrograde & symmetric & (9), (11--12) \\
        prograde  columnar   & thermal Rossby; Busse columns  & not yet & prograde & anti-symmetric  & (9), (14--18)\\\hline
    \end{tabular}
    \label{tab:modes}
    \tablebib{ 
(1)~\cite{Loptien2018NatAs}; 
(2)~\cite{Liang2019A&A};
(3)~\cite{Proxauf2020A&A};
(4)~\cite{Hanson2024PhFl};
(5)~\cite{Gizon2021A&A};
(6)~\cite{Hanson2022NatAs};
(7) \cite{Gizon2020A&A};
(8)~\cite{Fournier2022A&A};
(9)~\cite{Bekki2022A&A}; 
(10)~\cite{Triana2022ApJL};
(11)~\cite{Bhattacharya2023ApJs};
(12)~\cite{Jain2024ApJ};
(13) \cite{Bekki2024A&A};
       (14)~\cite{Roberts1968RSPTA};
    (15)~\cite{Busse1970JFM};
   (16)~\cite{Glatzmaier1981ApJS};
   (17)~\cite{Hindman2022ApJ};
   (18)~\cite{Hindman2023ApJ}.
   }
   \tablefoot{HFR stands for high-frequency retrograde.}
\end{table*}

\begin{table*}
\small
    \caption{Overview of model assumptions used in the literature to compute eigenmodes.}
    \centering
    \begin{tabular}{c c c c c c c}
    
    \hline\hline
        publications & model assumptions & density & differential  & buoyancy  & latitudinal  &  super-  \\
        & & stratification & rotation & effects & entropy gradient & adiabaticity \\
        \hline
        (5), (9) &  compressible 2.5D & solar-like  & solar-like & yes & yes & yes \\
        (4), (11), (19) & anelastic 2.5D  & solar-like & solar-like & yes & no & yes \\
         (10), (20--21) &  incompressible 2.5D & unstratified
    & no & no & no & no \\
        (12), (16), (22) &    {sound-proof cylindrical}& polytropic  & no & yes & no & yes  \\ 
        (5), (7--8) & incompressible 1.5D (surface) & ... & solar-like & no & no & no \\
         \hline
    \end{tabular}
    \label{tab:literature}
    \tablebib{The first 18 references are as listed in Table~\ref{tab:modes}. Additional references: 
    %
    (19)~\cite{Bhattacharya2024ApJ}; 
    %
    (20)~\cite{Rieutord1997JFM};
    (21)~\cite{Baruteau2013JFM};
    %
    (22)~\cite{Jain2023ApJ}.
    %
    }
    \tablefoot{
    The notation 2.5D refers to a set of independent 2D (latitude-radius) problems, one for each azimuthal wavenumber $m$. 
    The notation 1.5D refers to a set of independent 1D equations in latitude, one for each $m$.
    }
\end{table*}


\section{Introduction}
The Sun undergoes various oscillations through which we can probe deep into its interior. The pressure modes (p modes) are acoustic modes of oscillation restored by pressure and have a timescale of minutes \citep[e.g.][]{Christensen-Dalsgaard2002RvMP}. They have been successfully used to infer the Sun's large-scale flows,  internal differential rotation \citep[e.g.][]{Schou1998ApJ,  
Larson2018SoPh} and meridional circulation \citep[e.g.][]{Giles1997Natur, Gizon2020Sci}. 
{Modes in the inertial frequency range -- `inertial modes' for short -- are low-frequency modes of oscillation restored by the Coriolis force \citep[e.g.][]{Greenspan1968Book}. Their periods are comparable to the Sun's rotation period.


All observed inertial modes were first detected in the horizontal flow field on the Sun's surface. They are most easily described as waves of radial vorticity that propagate retrograde in a frame that co-rotates with the Sun's Carrington rotation rate (approximately the equatorial rotation rate). 
By convention, in this paper, we use the Carrington rotation rate as a reference when using the terms retrograde and prograde.
The observed inertial modes belong to several classes \citep[see][for a review]{Gizon2024IAU}. 
The equatorial Rossby modes follow the well-known classical dispersion relation for the sectoral Rossby modes of a uniformly rotating fluid for azimuthal wavenumbers in the range $3\leq m\lesssim 30$ \citep[][]{Loptien2018NatAs, Liang2019A&A, Proxauf2020A&A, Hanson2024PhFl}. 
Their retrograde propagation results from the planetary $\beta$~effect, which arises due to the latitudinal variation of the tangential component of the Coriolis force \citep[e.g.][]{Vallis2017book}. 
}
{\cite{Gizon2021A&A} found many additional quasi-toroidal global inertial modes of oscillation in the solar data.
Among these modes are high-latitude modes with $m=1,2,3$. These global modes can have either symmetric or antisymmetric north-south vorticity. Their velocity eigenfunctions display a characteristic spiral pattern in the polar regions.
The mode with the largest velocity amplitude ($\sim 15$~m/s) is the $m=1$ high-latitude mode with north-south symmetric radial vorticity; it is present in daily-cadence solar Dopplergrams over the last five solar cycles \citep{Liang2025A&A}.
We note that the velocity pattern associated with this $m=1$ mode had been reported before but misidentified as giant convection cells \citep{Hathaway2013Sci}. 
In addition to high-latitude modes, many other modes, with $m$ up to at least 10, have been observed with amplitudes that peak at mid-latitudes, near their critical latitudes. The critical latitude of a mode is the latitude where its phase speed equals the local rotation velocity. Modes that are retrograde (with respect to the equatorial rotation rate) can have critical latitudes since the Sun's rotation rate decreases with latitude (with the sharpest decrease above $60^\circ$). All the modes described above, including the equatorial Rossby and high-latitude modes, have critical latitudes.

Furthermore, \cite{Hanson2022NatAs} detected retrograde modes with equatorially-antisymmetric radial vorticity. These modes do not follow the dispersion relation of any classical Rossby modes. Since they propagate with phase speeds about three times that of the equatorial Rossby modes, \citet{Hanson2022NatAs} referred to them as high-frequency retrograde (HFR) modes. {Many \HFR\ modes also have critical latitudes, as they propagate in the retrograde direction.}


Under the assumption that the modes have small enough amplitudes, linear modelling is key to identifying the modes and revealing their physical nature.
It was recognized early on that solar latitudinal differential rotation and turbulent viscosity are crucial ingredients to describe the purely toroidal modes on the sphere \citep{Gizon2020A&A, Fournier2022A&A}. 
{
The frequency spectrum under differential rotation differs significantly from that under uniform rotation.}
\cite{Bekki2022A&A} also solved the eigenvalue problem in a spherical shell representative of the differentially rotating convection zone. 
The linear modes in 1.5D of  \citet{Fournier2022A&A} and 2.5D of \cite{Bekki2022A&A} were both essential in identifying the observed modes and providing a unified explanation for the equatorial Rossby modes and the mid- and high-latitude modes, including the $m=1$ high-latitude mode \citep[see][]{Gizon2021A&A, Gizon2024IAU}.
It was also found that the high-latitude modes are baroclinically unstable and sensitive to the latitudinal entropy gradient in the convection zone. 
Linear modelling has also been instrumental in identifying the \HFR\   modes as non-toroidal retrograde mixed inertial modes with anti-symmetric radial vorticity \citep{Triana2022ApJL, Bhattacharya2023ApJs, Bekki2024A&A, Jain2024ApJ}.

{Furthermore, some linear studies have described retrograde modes that are similar to the equatorial Rossby modes but with $n=1$, {where $n$ is the number of nodes in radius} \citep{Bekki2022A&A, Bhattacharya2023ApJs,  Jain2024ApJ}. }  {These modes have been referred to as $n=1$ equatorial Rossby modes \citep{Bekki2022A&A}. They also belong to the class of retrograde mixed modes, similar to the \HFR\ modes, but with north-south symmetric radial vorticity \citep{Bekki2022A&A, Jain2024ApJ}. In order to label this particular subset of mixed modes, we stick to the name \Rossbyno\ mode. These modes have frequencies comparable to the sectoral power of radial vorticity in the solar surface flows for $m\gtrsim 8$ \citep{Hanson2024PhFl}. However, the observed power cannot be conclusively attributed to these modes due to the lack of comparison between their eigenfunctions and the solar flow structures. Adding to that, the frequencies and growth rates of these linear modes are sensitive to the superadiabaticity, which is not well-constrained for the Sun \citep{Hanson2024PhFl, Bekki2022A&A}.
}

Linear calculations previously predicted modes that are excited by thermal instability in a rotating fluid \citep{Roberts1968RSPTA, Busse1970JFM}. These are non-toroidal modes consisting of convective columns that propagate in the prograde direction under the effect of compressional and topographical $\beta$~effects \citep[e.g.][]{Glatzmaier1981ApJS, Gastine2014PEPI, Bekki2022A&A, Hindman2022ApJ}. 
These modes are also referred to as columnar convective modes, Busse columns, banana cells, and thermal Rossby modes.
 In our paper, we refer to them as prograde columnar modes.  In non-linear numerical simulations, prograde columnar modes play a crucial role in the dynamics of the convection zone as they transport angular momentum equatorward and contribute to determining the differential rotation profile \citep[e.g.][]{Miesch2006ApJ}. 
 {They also transport heat poleward, which helps establish the thermal wind balance \citep[e.g.][]{Matilsky2020ApJ}. }
Despite their important function in simulations, they have not yet been identified in the solar surface flows. {
There are several possible reasons why they have not been observed.
{They may be present below the near-surface layers but shielded by small-scale convection close to the solar surface, or their spatial scale may be too small to be identified in the solar surface flows \citep{Hindman2022ApJ, Featherstone2016ApJ}.} Also, the modes may simply not be present at measurable amplitudes in the solar convection zone, which could be related to the unresolved convective conundrum 
\citep[e.g.][]{Hotta2023SSRv}. 
}

In our paper, we study the above-discussed modes as summarized in Table~\ref{tab:modes}. Understanding the physics of inertial modes can help diagnose different properties of the solar interior to complement p~mode helioseismology, as they are sensitive to several parameters of the deep convection zone, including the turbulent viscosity, superadiabaticity, and latitudinal entropy gradient \citep{Gizon2021A&A, Bekki2022A&A, Hanson2024PhFl}. More importantly, the baroclinically unstable high-latitude modes likely play a key role in the dynamics of the Sun by controlling the solar differential rotation \citep{Bekki2024SciA}.

\section{Previous theoretical studies:  {Various assorted assumptions}}

{
\cite{Rieutord1997JFM} numerically analysed the spectrum of inertial waves trapped in a spherical shell containing a uniformly rotating incompressible fluid. \cite{Baruteau2013JFM} extended this setup without stratification to a differentially rotating fluid. 
Following the discovery of solar {quasi-toroidal inertial modes} \citep{Loptien2018NatAs, Gizon2021A&A}, several attempts have been made to model the various classes of solar inertial modes employing varying simplifying assumptions. 
\cite{Bekki2022A&A} modelled the observed solar inertial modes using a differentially rotating compressible fluid in the solar convection zone. 
Furthermore, \cite{Bhattacharya2023ApJs, Bhattacharya2024ApJ} modelled the solar inertial modes in the spherical shell geometry considering the anelastic approximation \citep{Glatzmaier1981ApJS}. This approximation assumes that the sound speed is much faster than the other characteristic flow speeds in the system, thereby filtering out the acoustic modes. 
 {Although these sound-proof setups often do not conserve energy for gravity waves in the sub-adiabatic radiative interior, they generally conserve energy in the nearly adiabatic convection zone \citep{Brown2012ApJ}. 
}
\cite{Triana2022ApJL} also used a simpler uniformly rotating unstratified incompressible fluid in a spherical shell to model the solar inertial modes. The study aimed to identify the \HFR\ modes previously discovered by \cite{Hanson2022NatAs}. }
Such an incompressible model is widely used to study planetary eigenmodes employing the Boussinesq approximation to include the buoyancy effects \citep[e.g.][]{Marti2016GGG, Barik2023E&SS}.
Furthermore, several studies have been conducted in simplified geometries, such as on spherical surfaces \citep{Fournier2022A&A, Gizon2020A&A} and in cylindrical geometry \citep{Jain2023ApJ, Jain2024ApJ}.  {We refer to the model of \cite{Jain2023ApJ} as sound-proof because it assumes the limit of low frequency and high sound speed to filter out the sound waves.}

{
{The setups in these linear models for the solar inertial modes differ in various aspects}, such as the numerical domain,  the degree of superadiabaticity, the choice of background rotation profile, and the latitudinal entropy gradient in the convection zone.
 Many studies, including \cite{Triana2022ApJL}, \cite{ Bhattacharya2023ApJs}, \cite{Jain2023ApJ}, and \cite{Jain2024ApJ} assumed solid-body rotation for the Sun. However, as discussed earlier, latitudinal differential rotation implies the existence of critical latitudes and additional families of modes (e.g. the high-latitude modes). }
Table~\ref{tab:literature} summarizes the assorted fundamental model assumptions in the various linear simulations of the solar inertial modes.
Although the solar inertial modes have also been studied using nonlinear simulations \citep[e.g.][]{Bekki2022A&A2, Bekki2024SciA, Blume2024ApJ}, we limit our studies to the linear modes in the solar convection zone. 

Despite employing several differing simplifications, the effects of these simplifying assumptions on the linear inertial modes have not been tested. 
\cite{Bekki2024A&A} compared only the \HFR\ modes obtained from the different models used in the literature employing spherical shell geometry. 
 {The effects of various sound-proof approximations were tested earlier for gravity waves in stellar interiors \citep{Brown2012ApJ, Vasil2013ApJ}, but not for inertial modes. }

}

\section{Description of our numerical computations: Compressible, anelastic, and Boussinesq cases}\label{Model_descriptions}

In this work, {we study the effects of the anelastic and Boussinesq approximations on the solar inertial modes by comparison to the fully compressible case used as the reference. 
We implement the compressible, anelastic, and Boussinesq model setups in Dedalus \citep{Burns2020PRR} with the same boundary conditions (stress-free impenetrable). 
The background stratification is the same for the compressible and the anelastic cases. 
The background temperature, superadiabaticity, and latitudinal entropy variations are identical in all computations, including the Boussinesq case. 
This helps us understand the main physical ingredients needed to describe the inertial modes correctly.} 
We note that we always represent the effects of small-scale turbulence in terms of turbulent eddy viscosity. 
The effects of magnetic fields on the inertial modes are ignored in our study. 
{ We discuss the five types of inertial modes described in Table~\ref{tab:modes}.}

\begin{figure}
    \centering
    \includegraphics[width=0.95\linewidth]{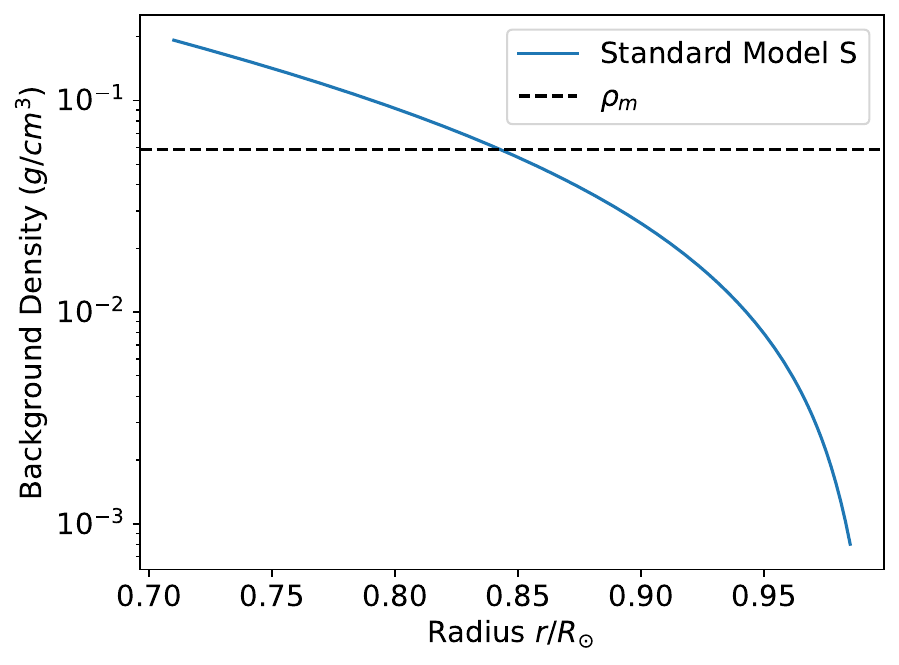}
    \caption{Variation of the background density in the solar convection zone from the standard model S \citep{Christensen-Dalsgaard1996Sci}, denoted by the blue curve. The black dashed line shows the mean density $\rho_m$ in the solar convection zone. The Boussinesq setup uses the mean density, whereas the other stratified setups use the density from model S.
    }
    \label{fig:1}
\end{figure}

\subsection{Linearized equations of fluid dynamics}\label{equations}
The linearized Navier-Stokes equation in the Carrington frame, which rotates at $\Omega_0/2\pi=456 $ nHz, can be expressed as
\begin{multline}\label{eq:1}
    \frac{\partial \textbf{u}}{\partial t}+\textbf{v}_0\cdot\nabla \textbf{u}+\textbf{u}\cdot\nabla \textbf{v}_0+2 \Omega_0\,\textbf{e}_z\times \textbf{u}\\
    +\mathbf{\nabla}\left(\frac{p_1}{\rho_0}\right) -\frac{s_1}{c_p}g\,\textbf{e}_r  - \frac{1}{ \rho_0}  \nabla\cdot \mathbf{\overleftrightarrow{D}}=0.
\end{multline}
Here, $\textbf{u}$, $\rho_1$, $p_1$, and $s_1$ denote the perturbations of velocity, density, pressure, and entropy with respect to the background, while $c_p$ is the specific heat at constant pressure. The viscous stress tensor  is given by
$\overleftrightarrow{D}=\rho_0\nu \left( \nabla \textbf{u} + \nabla \textbf{u}^T - \frac{2}{3}(\nabla \cdot \textbf{u}) \overleftrightarrow{I} \right),$
with $\nu=10^{12} \, \rm cm^2/s$ being the spatially constant isotropic turbulent viscosity and $ \overleftrightarrow{I}$ the identity tensor.
The background flow $\textbf{v}_0=(\Omega(r,\theta)-\Omega_0)\, r \sin \theta \, \textbf{e}_{\phi}$ represents the solar differential rotation $\Omega(r,\theta)$  in the convection zone taken from \citet{Larson2018SoPh}. In this study, we omit the effects of meridional circulation, which has been shown to have a small impact \citep{Gizon2020A&A, Fournier2022A&A}.
For compressible and anelastic setups, the background stratifications of density $\rho_0$, pressure $p_0$, and temperature $T_0$ are taken from the standard solar model~S \citep{Christensen-Dalsgaard1996Sci}. The acceleration due to gravity $g$ is determined using the hydrostatic balance in model S. 
 {For the anelastic setup, Eq.~\eqref{eq:1} is the form of the Navier-Stokes equation that is recommended by \cite{Brown2012ApJ}, as it conserves energy both in the convection zone and the radiative zone.  
}
For the incompressible setup, we apply the Boussinesq approximation following \cite{Spiegel1960ApJ} to account for buoyancy effects, ensuring similar conditions to the other models and allowing baroclinic instability \citep[e.g.][]{Molemaker2005JPO}.
This is in contrast to the model used in \cite{Triana2022ApJL},  which ignored the buoyancy effects.
Hence, we refer to it as the Boussinesq model instead of the incompressible model in our study. Under this approximation, we use the mean density, {$\rho_m = 0.0589$~g~cm$^{-3}$}, in the convection zone as the background density $\rho_0$. Figure~\ref{fig:1} depicts the difference between the mean density and the background density from model S. However, the temperature $T_0$ is used from the Solar model S. The background pressure $p_0$ is given by the hydrostatic equilibrium, similar to \cite{Spiegel1960ApJ}. See Appendix~\ref{appendix:equations} for more details.

The following linearized equation of entropy ($s$) is used
\begin{equation}\label{eq:2}
    \frac{\partial s_1}{\partial t} + \textbf{v}_0\cdot\nabla s_1 + u_r\frac{\partial s_0}{\partial r} + \frac{u_{\theta}}{r} \frac{\partial s_0}{\partial \theta} - \frac{1}{ \rho_0 T_0}\nabla \cdot ( \kappa \rho_0 T_0 \nabla s_1)=0,
\end{equation}
where $s_0$ denotes the background entropy. In this study, we assume the adiabatic stratification in radius, ${\partial s_0}/{\partial r}=0$, for simplicity. The latitudinal entropy gradient $\partial s_0/\partial\theta$ represents the thermal wind balance of the solar differential rotation and is estimated as \citep[e.g.][]{Pedlosky1982book, Thompson2003ARA&A}
\begin{equation}
    \frac{\partial s_0}{\partial\theta} = r^2 \sin{\theta} \frac{g}{c_p}  \frac{\partial (\Omega^2)}{\partial z},
\end{equation}
where $z$ is the coordinate along the rotational axis.
The turbulent thermal diffusivity $\kappa$ is assumed to be spatially constant, with a value equal to $\nu$.

For a compressible fluid, the linearized continuity equation
\begin{equation}\label{eq:4}
    \frac{\partial \rho_1}{\partial t}+\mathbf{\nabla}\cdot(\rho_0 \textbf{u})+\mathbf{\nabla}\cdot(\rho_1 \textbf{v}_0)=0, \ \ \ \mathrm{(compressible)} 
\end{equation}
is used in combination with the linearized equation of state
\begin{equation}
    \frac{p_1}{p_0}=\gamma\frac{\rho_1}{\rho_0}+\frac{s_1}{c_v},
\end{equation}
where $\gamma=5/3$ is the specific heat ratio, and $c_v$ is the specific heat at constant volume.
Under the anelastic approximation, we assume that the sound speed is much faster than the rotational or advective flow speed in the Sun. Thus, the continuity equation is reduced to the anelastic equation, 
\begin{equation}\label{eq:anelastic}
    \mathbf{\nabla}\cdot(\rho_0 \textbf{u})=0. \ \ \ \mathrm{(anelastic)} 
\end{equation} 
Under the Boussinesq approximation, this is further reduced to 
\begin{equation}\label{eq:boussinesq}
   \nabla\cdot \textbf{u}=0. \ \ \ \mathrm{(Boussinesq)} 
\end{equation} 
The above linearized fluid dynamical equations are used in this work to model and understand the physics of solar inertial modes.

\subsection{Boundary conditions}
{The Sun does not have any hard boundaries, but to execute the computations, one needs to implement some boundary conditions. As in some previous linear studies, we perform our computations in the spherical shell of the solar convection zone. The numerical domain in our study extends from the base of the convection zone ($r_i=0.71R_{\odot}$) to slightly below the photosphere ($r_o=0.985R_{\odot}$). There are various possible options for the boundary conditions, such as free-surface boundary conditions, stress-free boundary conditions, and no-slip boundary conditions. It is not very clear which boundary condition would be the best for modelling the solar inertial modes. The no-slip boundary condition is not expected to be good in modelling the inertial modes as it does not allow any motions at the boundaries. The free-surface boundary condition might be adequate to model the inertial modes as it allows a flexible boundary height. However, to avoid the computational complexity, we use the stress-free impenetrable boundary conditions, which have been used in previous studies of the solar inertial modes  \citep[e.g.][]{Bekki2022A&A, Bhattacharya2024ApJ, Triana2022ApJL}. This works better than the no-slip boundary conditions as it allows horizontal motions at the boundaries, although it assumes no vertical motions at the boundaries. Also, we assume no flux of entropy across each boundary. The boundary conditions can be expressed as 
\begin{align}\label{eq:8}
  &\   u_r(r_i)= u_r(r_o)=0,\\
&\overleftrightarrow{D}_{r\theta} (r_i)=\overleftrightarrow{D}_{r\theta} (r_o)={0},\\ 
&\overleftrightarrow{D}_{r\phi}(r_i)=\overleftrightarrow{D}_{r\phi}(r_{o})={0},\\ 
&\left. \frac{\partial s_1}{\partial r} \right|_{r_i}=\left. \frac{\partial s_1}{\partial r} \right|_{r_o}=0.\label{eq:11}
\end{align}
Since we use spherical harmonics as basis functions, the problem has no singularities at the poles and does not require any boundary conditions there \citep{Boyd2001Book}.
}

\begin{figure*}
    \centering
    \includegraphics[width=0.9\textwidth]{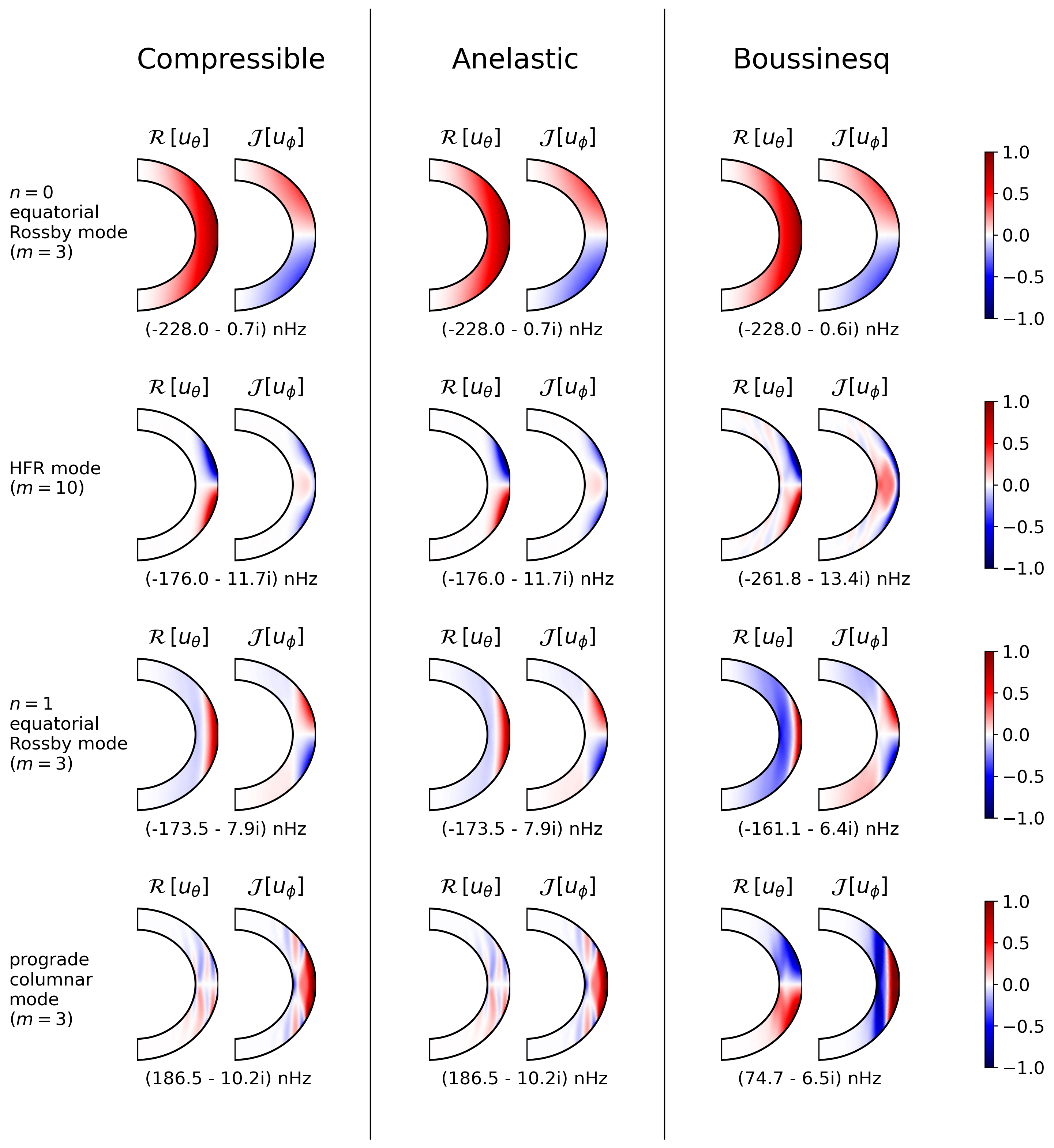}
    \caption{{Comparison of the eigenmodes of the different classes of inertial modes (see Table~\ref{tab:modes}) computed using compressible, anelastic, and Boussinesq models under uniform rotation. Note that the unstable high-latitude inertial modes are absent under solid body rotation. Here, we plot the real part of $u_\theta$ and the imaginary part of $u_\phi$ of the computed eigenmodes. The longitudes corresponding to the real and imaginary phases of the eigenfunctions are $\phi=\phi_0$ and $\phi=\phi_0-\pi/2m$, respectively, where $\phi_0$ is a longitude, where $u_{\theta}$ attains its maximum. The corresponding frequencies measured in the Carrington frame are stated below each eigenmode. The imaginary parts of the frequencies indicate the growth rates of the modes. All eigenfunctions are normalized such that the maximum of $u_\theta$ is 1 m/s at the surface.} }
    \label{fig:2}
\end{figure*}

\subsection{Formulation of eigenvalue problem in Dedalus}
{We use Dedalus \citep{Burns2020PRR}, a flexible open-source spectral code, to solve the linear eigenvalue problem of the solar inertial modes.
The above Eqs.~\eqref{eq:1}, \eqref{eq:2}, and \eqref{eq:4} -- \eqref{eq:boussinesq} are solved as an eigenvalue problem using the wave Ansatz where each perturbed physical quantity is proportional to $ \exp({\text{i} m\phi-\text{i}\omega t})$, with $m$ the azimuthal order and $\omega$ the frequency. We implement Eqs.~\eqref{eq:8} -- \eqref{eq:11} as the boundary conditions.} The calculations are performed on the spherical shell basis of Dedalus using the sparse eigenvalue solver \citep{Burns2020PRR}. We solve the sparse problem spanning the range of the inertial frequencies, $|\Re[\omega]| \leq 2\Omega_0$.
This helps us filter out the low-frequency modes of our interest. 
The obtained eigenfrequencies $\omega$ and eigenfunctions of the inertial modes are analysed for different model assumptions. All the eigenmodes we present in this paper are obtained using a grid on the spherical shell basis with 24 radial and 124 latitudinal points. We detect them based on their north-south symmetries, growth rates, and other known properties, such as the number of radial or latitudinal nodes. The convergence errors in the eigenfrequencies obtained by halving or doubling the resolutions are orders of magnitude smaller than the frequency resolution of observations \citep[which is on the order of a few nHz, see][]{Gizon2021A&A}. The reported eigenfunctions are also well-converged.

\begin{figure}
    \centering
    \includegraphics[width=0.49\textwidth]{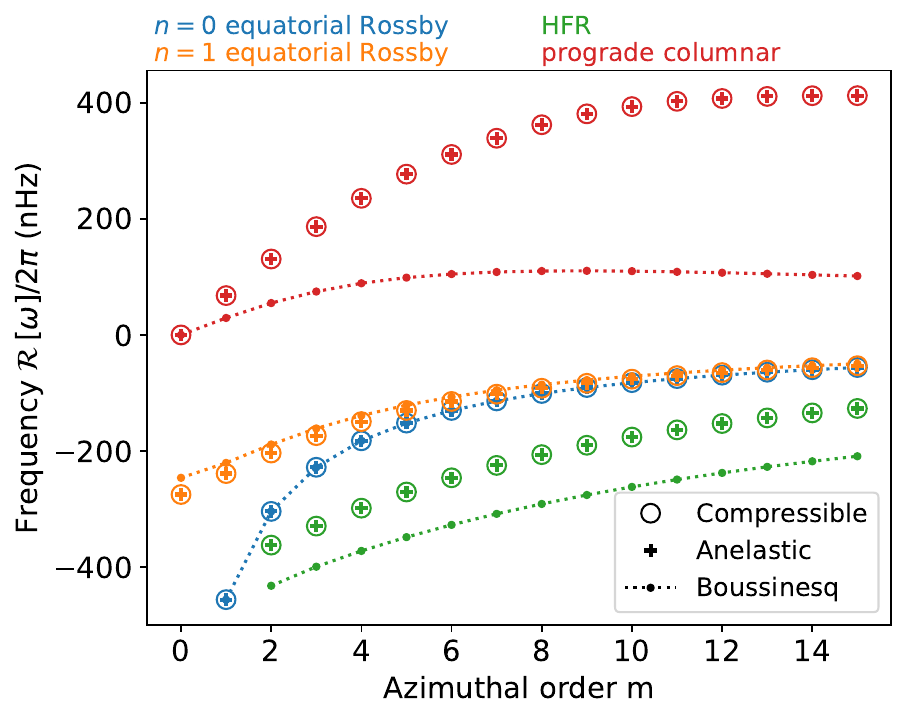}
    \caption{
    {Dispersion relations of the studied classes of inertial modes under uniform rotation for the range of azimuthal orders $0 \leq m \leq 15$. The blue, green, orange, and red colours represent the \Rossby, the \HFR, the \Rossbyno, and the prograde columnar modes, respectively. Open circles, cross symbols, and dotted points show the results computed from the fully compressible, anelastic, and Boussinesq models, respectively.}
    }
    \label{fig:3}
\end{figure}

\begin{figure}
    \centering
    \includegraphics[width=0.45\textwidth]{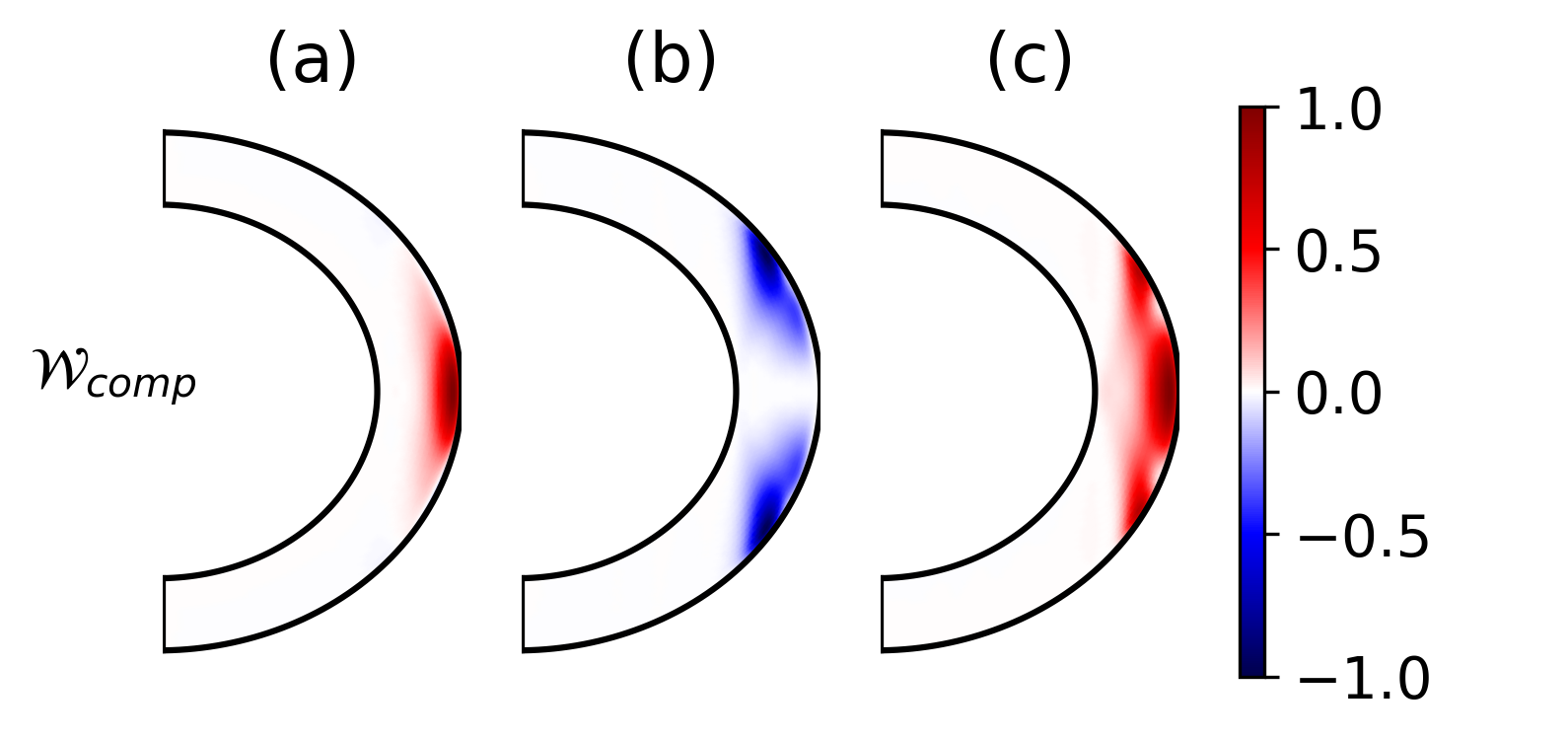}
    \caption{
    {Estimates of the importance of the compressional $\beta$~effect under uniform rotation using $\Re[ \mathcal{W}_{\rm comp}]$, as defined in Eq.~\eqref{def-W}, for the compressible setup. A negative value of $\Re[ \mathcal{W}_{\rm comp}]$ implies that the compressional $\beta$~effect promotes retrograde propagation, while a positive value implies that it promotes prograde propagation. Column (a) denotes the \HFR\ mode with $m=10$, column (b) denotes the \Rossbyno\ mode with $m=3$, and column (c) denotes the prograde columnar mode with $m=3$. For all modes, we normalize  $\Re[ \mathcal{W}_{\rm comp}]$ such that its maximum absolute value is 1.}
    }
    \label{fig:4}
\end{figure}

\section{Effects of simplifying the continuity equation}

\subsection{Effects of anelastic and Boussinesq approximations under uniform rotation} \label{sec:uniform}
We start by comparing the eigenmodes obtained from the different models under uniform rotation. Figure~\ref{fig:2} compares the eigenfunctions of the \Rossby\ mode with $m=3$,  the \HFR\ mode with $m=10$, the \Rossbyno\ mode with $m=3$, and the prograde columnar mode with $m=3$ for the three different models. First, we note that the eigenmodes in the compressible and anelastic models are nearly identical. They have negligible differences in the eigenfunctions of less than  1\%, and their frequencies are the same correct up to 0.1~nHz. However, the eigenmodes in the Boussinesq model differ significantly from those obtained using the other models, except for the \Rossby\ modes.
In all the models, the \Rossby\ modes are almost perfectly toroidal and are driven by the planetary $\beta$~effect, which is unaffected by the background density stratification.
However, the other modes, having a substantial non-toroidal nature, are quite different in the Boussinesq model as compared to the other models.
Significant differences observed in the spatial eigenfunctions of the \HFR\ modes, the \Rossbyno\ modes, and the prograde columnar modes can be understood as follows:
In contrast to the \Rossby\ modes, these non-toroidal modes exhibit a radial node in the horizontal velocity eigenfunctions within the convection zone (CZ).
Under the background density stratification, the constraint of local mass conservation (Eq.~\ref{eq:anelastic}) requires the mass fluxes in the lower and upper parts of CZ (across the nodal plane) to be balanced with each other, leading to slower velocities in the lower CZ.
In the Boussinesq model, on the other hand, the velocity eigenfunctions can have comparable amplitudes throughout the CZ.

Figure~\ref{fig:3} presents the dispersion relations of the modes displayed in Fig.~\ref{fig:2} for the three models under uniform rotation, covering azimuthal orders $m$ from 0 to 15. Similar to the eigenmodes, the dispersion relations of the various modes are identical between the anelastic and compressible models across all values of $m$.  
The deviations in the dispersion relations of the Boussinesq model relative to the other models are evident for the non-toroidal modes discussed above. The deviations are enormous for the prograde columnar modes and the \HFR\ modes. In both cases, the frequencies are significantly shifted to the negative (more retrograde) direction. Due to their less toroidal nature, the deviations are smaller for the \Rossbyno\ modes. The purely toroidal \Rossby\ modes have the same dispersion relations for all the models.

To investigate the effects of using the Boussinesq approximation on mode frequencies, we examine the linearized vorticity equation.
By taking a curl of Eq.~\eqref{eq:1} under uniform rotation, we express the radial and $z$ components of the vorticity equation as
\begin{eqnarray}
&&    \frac{\partial {\zeta_r}}{\partial t} \approx \underbrace{\frac{2\Omega_0 \sin \theta }{r}u_\theta}_{\mathrm{planetary }\ \beta\ \mathrm{effect}}  - \frac{2\Omega_0 \cos \theta}{H_\rho} u_r  \nonumber \\
&&   \ \ \ \ \ \  \ \ \ \ \ \  \ \ \ \ \ \  \ \ \ \ \ \    + 2\Omega_0 \cos \theta  \frac{\partial u_r}{\partial r} - \frac{2\Omega_0 \sin \theta }{r} \frac{\partial u_r}{\partial \theta}, \label{eq:radial-vorticity1} \\
&& \frac{\partial {\zeta}_z}{\partial t} \approx \underbrace{- \frac{2\Omega_0}{H_\rho}u_r}_{\mathrm{compressional}\ \beta\ \mathrm{effect}}  + \underbrace{2\Omega_0   \frac{\partial u_z}{\partial z}}_{\mathrm{topographic}\ \beta\ \mathrm{effect}} -\frac{g}{c_p r} \frac{\partial s_1}{\partial \phi}, \label{eq:z-vorticity1}
\end{eqnarray}
where $\bm{\zeta}=\nabla \times \textbf{u}$ and $H_\rho$ is the density scale height. For simplicity, we omit the viscous diffusive terms.
Here, the first term on the right-hand side of Eq.~(\ref{eq:radial-vorticity1}) represents the planetary $\beta$~effect, while the first term on the right-hand side of the Eq.~(\ref{eq:z-vorticity1}) represents the compressional $\beta$~effect.
The second term on the right-hand side of the Eq.~(\ref{eq:z-vorticity1}) corresponds to the topographic $\beta$~effect (when integrated over $z$).
To assess the relative importance of these $\beta$~effects on the mode frequencies, we further transform the above equations into the following form:
\begin{eqnarray}
   &&  \omega |{\zeta}_r|^2 \approx \mathcal{G}_{\rm planetary} + \mathcal{G}_{\rm comp} + \mathcal{G}_{\rm other}, \label{eq:r-vorticity-final} \\
   &&  \omega |{\zeta}_z|^2 \approx \mathcal{W}_{\rm comp} + \mathcal{W}_{\rm topographic} + \mathcal{W}_{\rm other}, \label{eq:z-vorticity-final}
\end{eqnarray}
with
\begin{eqnarray}\label{beta-G}
&&    \mathcal{G}_{\rm planetary}=i \frac{2\Omega_0 \sin \theta }{r}u_\theta \zeta_r^*, \\
&&    \mathcal{G}_{\rm comp}=- i \frac{2\Omega_0 \cos \theta}{H_\rho} u_r \zeta_r^*, \\
&&    \mathcal{G}_{\rm other}=2 i\Omega_0 \cos \theta  \frac{\partial u_r}{\partial r} \zeta_r^* - i\frac{2\Omega_0 \sin \theta }{r} \frac{\partial u_r}{\partial \theta}\zeta_r^*,
\end{eqnarray}
and
\begin{eqnarray}\label{def-W}
&&    \mathcal{W}_{\rm comp}=-i \frac{2\Omega_0}{H_\rho} u_r\zeta_z^*, \\
&&    \mathcal{W}_{\rm topographic}= 2i\Omega_0   \frac{\partial u_z}{\partial z}\zeta_z^*, \\
&&    \mathcal{W}_{\rm other}=\frac{ m\, g}{c_p r}s_1\zeta_z^*. \label{beta-W}
\end{eqnarray}
{Here, $^*$ denotes the complex conjugate of the quantities. A negative real part of $\mathcal{G}$ or $\mathcal{W}$ indicates that the associated physical effect promotes retrograde propagation, whereas a positive real part implies that it promotes prograde propagation.}
Figures~\ref{fig:8} -- \ref{fig:11} present all terms in Eqs.~(\ref{eq:r-vorticity-final}) and (\ref{eq:z-vorticity-final}) for the \Rossby\ mode ($m=3$), the \HFR\ mode ($m=10$), the \Rossbyno\ mode ($m=3$), and the prograde columnar mode ($m=3$) from compressible and Boussinesq setups. Note that $\mathcal{G}_{\rm comp}=\mathcal{W}_{\rm comp}=0$ in the Boussinesq setup because the assumption of constant density leads to an infinite density scale height $H_\rho$.

Our analysis reveals that the primary cause of frequency changes from the anelastic (or compressible) to Boussinesq models is the absence of $\mathcal{W}_{\rm comp}$ in the $z$-vorticity equation.
Figure~\ref{fig:4} presents $\Re[\mathcal{W}_{\rm comp}]$ in the compressible setup for the \HFR\ mode with $m=10$, \Rossbyno\ mode with  $m=3$, and the prograde columnar mode with $m=3$.
As for the \Rossbyno\ modes, $\Re[\mathcal{W}_{\rm comp}]$ is negative, suggesting that the compressional $\beta$~effect promotes their retrograde propagation. This occurs because, in these modes, the radial vortical motions are dominant and $z$-vorticity $\zeta_z$ is primarily generated by the strong radial shear of longitudinal flows and is not associated with their radial motions ($\zeta_z$ and $-\partial u_r/\partial\phi$ have the opposite sign).
The absence of $\mathcal{W}_{\rm comp}$ in the Boussinesq model slightly shifts their mode frequencies towards the positive (more prograde) direction, as seen in Fig.~\ref{fig:3} for $m \leq 4$.
On the other hand, both \HFR\ modes and prograde columnar modes have strongly positive $\mathcal{W}_{\rm comp}$, indicating that the compressional $\beta$~effect enforces their prograde propagation.
This is because their radial motions are strongly associated with their $z$-vortices ($\zeta_z$ and $-\partial u_r/\partial\phi$ have the same sign).
Without the compressional $\beta$~effect, the frequencies of the \HFR\ and prograde columnar modes shift towards the negative (more retrograde) direction, as seen in Fig.~\ref{fig:3}.
We also note that, in prograde columnar modes, the planetary $\beta$~effect plays an additional role in decreasing their prograde frequencies in the Boussinesq model (see Fig.~\ref{fig:11}).
In the absence of density stratification, upflows converge towards the equator due to spherical curvature, enhancing the planetary $\beta$~effect.
This does not occur in compressible or anelastic models because the upflows tend to expand and drive horizontally diverging motions due to the background density stratification.

\begin{figure}
    \centering
    \includegraphics[width=0.49\textwidth]{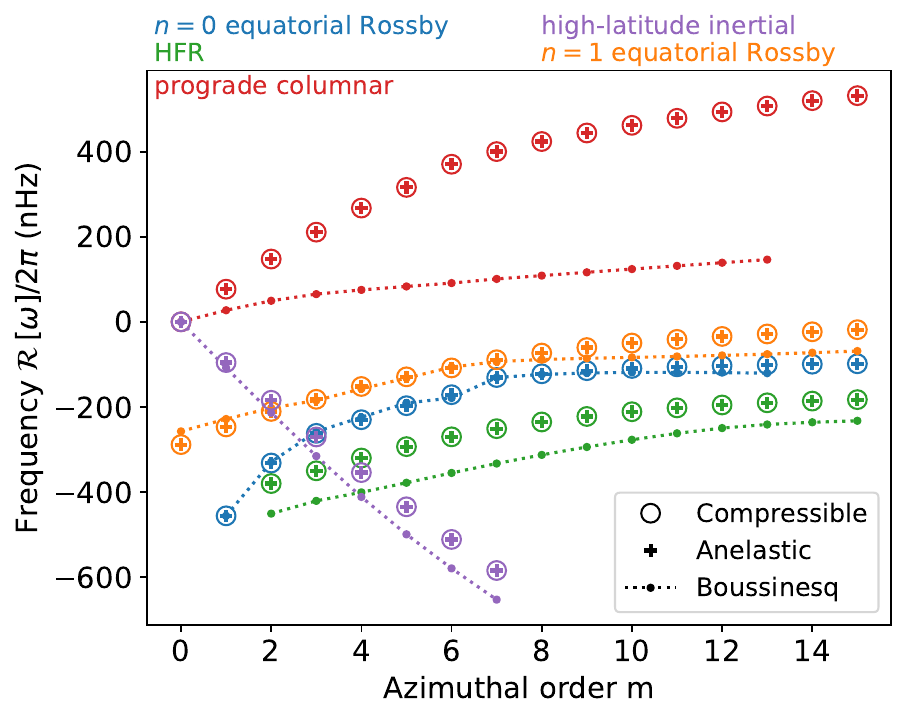}
    \caption{{Dispersion relations of the various classes of inertial modes listed in Table~\ref{tab:modes}, computed using different models with solar differential rotation and the associated latitudinal entropy gradient. The same notation applies to colours and symbols as in Fig.~\ref{fig:3}. The only addendum is the high-latitude mode with north-south symmetric radial vorticity, represented in purple. }
    }
    \label{fig:5}
\end{figure}

\begin{figure*}
    \centering
    \includegraphics[width=0.96\linewidth]{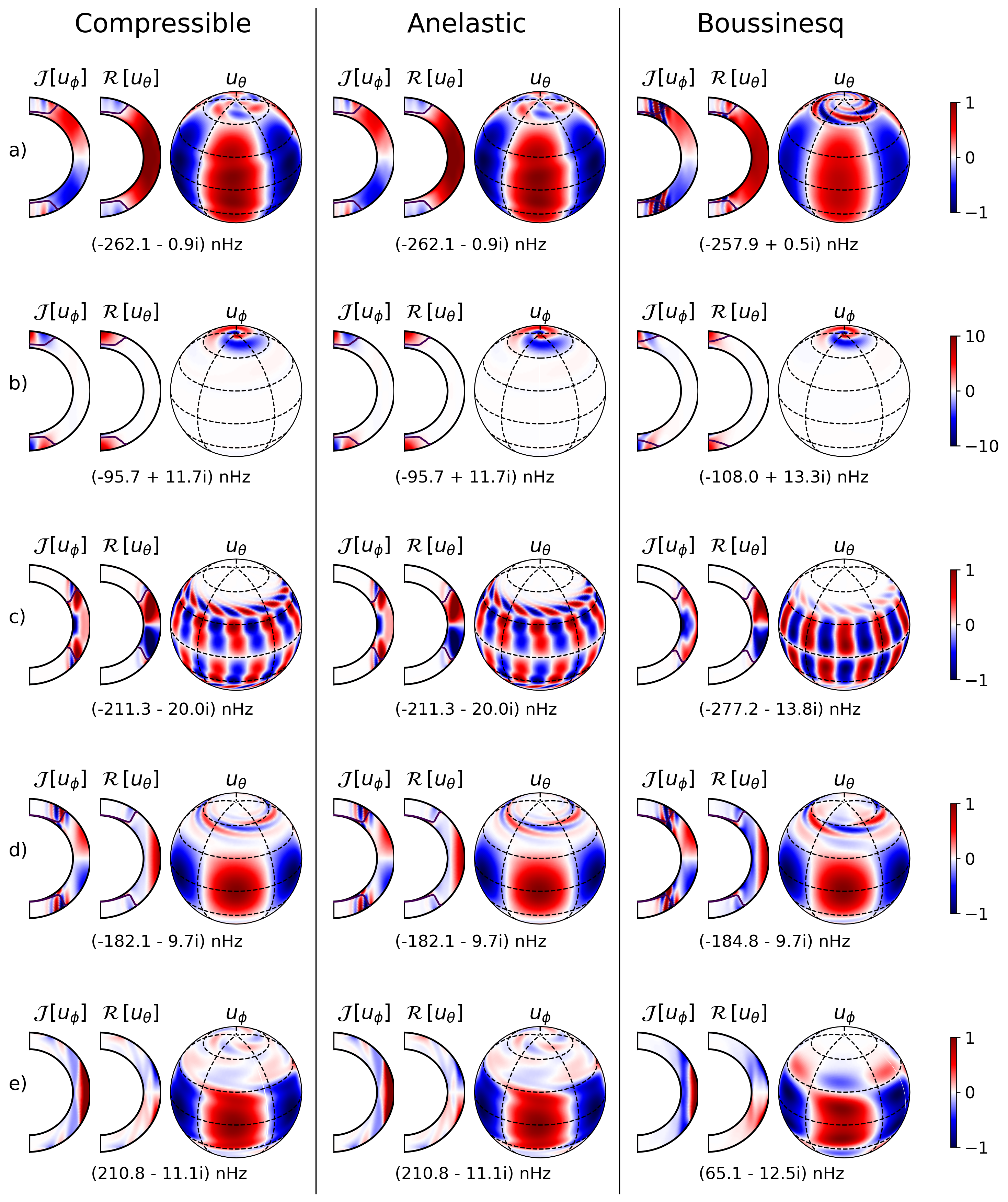}
\caption{{Comparison of the eigenmodes for different classes of inertial modes computed using compressible, anelastic, and Boussinesq models under solar differential rotation and the associated latitudinal entropy gradient. The different rows denote: a)~\Rossby\ mode ($m=3$), b)~high-latitude mode with north-south symmetric radial vorticity ($m=1$), c)~\HFR\ mode (${m=10}$), d)~\Rossbyno\ mode (${m=3}$), e)~prograde columnar mode (${m=3}$). Here, we plot the real part of $u_\theta$ and the imaginary part of $u_\phi$ of the computed eigenmodes in the meridional plane. The corresponding longitudes are chosen in the same way as in Fig.~\ref{fig:2}. The black solid curves on the meridional cross-sections denote the critical latitudes where $\Re [\omega]=m(\Omega-\Omega_0)$. We also show the surface velocity: $u_\phi$ for the high-latitude mode and the prograde columnar mode, and $u_\theta$ for the other modes. The corresponding frequencies in the Carrington frame are stated below each eigenmode. The imaginary parts of the frequencies are the growth rates of the modes. {All eigenfunctions, except for the high-latitude mode, are normalized such that the maximum of $u_\theta$ is 1 m/s at the surface. For the high-latitude mode, the maximum velocity at the surface is set to 10 m/s.} }}
    \label{fig:6}
\end{figure*}

\begin{figure}
    \centering
    \includegraphics[width=0.49\textwidth]{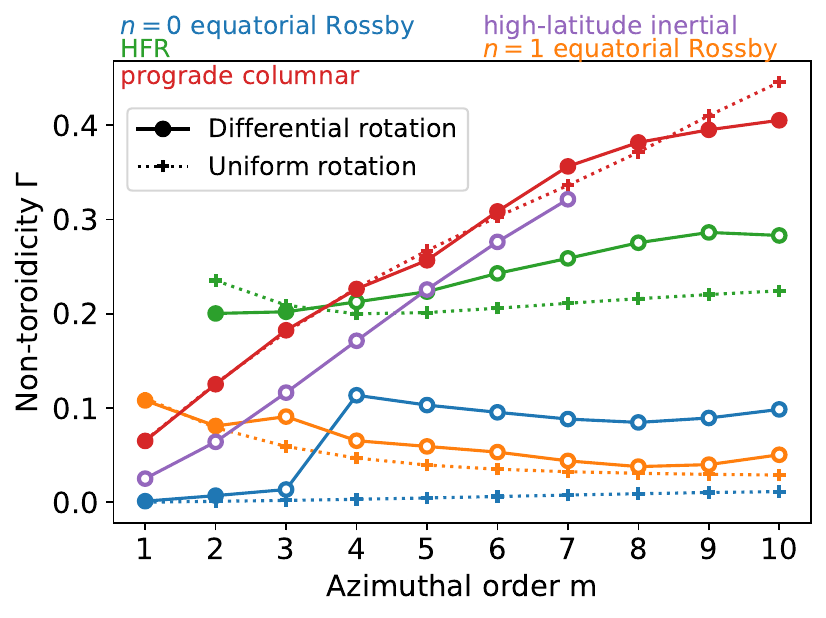}
    \caption{
    Estimates of the significance of the radial motions associated with the different inertial modes for the compressible setup, quantified using non-toroidicity $\Gamma$ defined in Eq.~\eqref{Non-toroidicity}. The values of non-toroidicity for different modes are plotted against azimuthal orders ranging from $m=1$ to $m=10$. Different colours represent different modes, as in Fig.~\ref{fig:5}. Solid lines denote modes under differential rotation, while dotted lines denote modes under uniform rotation. Open circles indicate modes affected by critical latitudes under differential rotation, while filled circles indicate other modes under differential rotation.
    }
    \label{fig:7}
\end{figure}

\subsection{Effects of anelastic and Boussinesq approximations under solar differential rotation and latitudinal entropy gradient}

Now, we include solar differential rotation and its associated latitudinal entropy gradient in our computations. Figure~\ref{fig:5} presents the dispersion relations of the different modes with $0\leq m\leq 15$ under this setup for all models. 
Here, we also consider the high-latitude modes, which are baroclinically unstable and owe their existence to the Sun's differential rotation and the latitudinal entropy gradient \citep{Bekki2022A&A}. {For simplicity, we only study the fastest-growing high-latitude mode for each azimuthal order. }
As in the uniform rotation case, the compressible and anelastic models yield the same dispersion relations for the different modes.
In the Boussinesq model, the frequencies of the prograde columnar modes and the \HFR\ modes shift in the negative (more retrograde) direction. In contrast, the frequencies of the \Rossby\ and \Rossbyno\ modes experience only marginal changes in the Boussinesq model.
These results are consistent with the uniform rotation case (\S~\ref{sec:uniform}).
The retrograde-propagating high-latitude modes exhibit a trend of negative frequency shift in the Boussinesq model, similar to that of the \HFR\ modes.
This arises from the absence of the compressional $\beta$~effect in the Boussinesq model, as high-latitude modes are essentially non-toroidal.

Figure~\ref{fig:6} compares the velocity eigenfunctions of the \Rossby\ mode with $m=3$, the high-latitude mode with $m=1$, the \HFR\ mode with $m=10$, the \Rossbyno\ mode with $m=3$, and the prograde columnar mode with $m=3$ in the differentially rotating case across the three models. 
As before, the eigenmodes obtained from the compressible and anelastic models are identical.
In contrast to the uniform rotation case, the eigenfunctions in the Boussinesq model differ even for the \Rossby\ modes under solar differential rotation.
We note that under differential rotation $\Delta\Omega(r,\theta)=\Omega(r,\theta)-\Omega_{0}$, the modes have critical latitudes $\theta_c$, where their phase speeds match the local differential rotation speed (i.e.~${\Delta\Omega(r,\theta_c)=\Re[\omega]/m}$).
Black solid curves in Fig.~\ref{fig:6} indicate the locations of the critical latitudes.
Previous studies \citep[][]{Gizon2020A&A, Fournier2022A&A} have shown that the mode eigenfunctions undergo strong distortion near the critical latitudes.
\Rossby, \HFR, and \Rossbyno\ modes are confined to the equatorial region by the critical latitudes. In contrast, the high-latitude modes exist at latitudes above the critical latitudes.
The columnar modes propagate prograde and lack critical latitudes.

Under differential rotation, substantial radial motions develop near the critical latitudes \citep{Bekki2022A&A}.
{To estimate the relative impact of the radial motions on the eigenmodes}, we define the non-toroidicity $\Gamma$ based on the ratio of the kinetic energy in the radial direction to the total kinetic energy of the mode:
\begin{equation}\label{Non-toroidicity}
     \Gamma =\sqrt{ \frac{  \int_{CZ} \rho_0 u_r^2 dV}{\int_{CZ} \rho_0( u_r^2 +u_{\theta}^2+ u_{\phi}^2) dV}}.
\end{equation}
Figure~\ref{fig:7} presents the non-toroidicity $\Gamma$ of the various modes for different azimuthal orders $m$.
Compared to uniform rotation, the non-toroidicity $\Gamma$ increases significantly due to critical latitudes under differential rotation.
This is most significant for the \Rossby\ modes at $m \geq 4$ but is also seen in the \HFR\ modes ($m \geq 4$) and the \Rossbyno\ modes~($m \geq 3$).
For prograde columnar and high-latitude modes, the non-toroidicity $\Gamma$ increases almost linearly with $m$.
Unlike the prograde columnar modes, the retrograde high-latitude modes are strongly influenced by the critical latitudes.
In the Boussinesq model, these modes exhibit more retrograde frequencies due to the absence of the compressional $\beta$~effect. Consequently, their critical latitudes shift polewards, confining the mode power to higher latitudes than in the compressible model (see Fig.~\ref{fig:6}).

\section{Summary and outlook}
In this work, we use a fixed framework in Dedalus to assess the validity of the anelastic and Boussinesq approximations for modelling the inertial modes in the solar convection zone. Firstly, we find that the eigenmodes obtained by the compressible and anelastic models are almost identical, regardless of the presence of the background solar differential rotation. The anelastic assumption does not affect the properties of the solar inertial modes since a scale separation between low-frequency inertial modes and very high-frequency sound waves occurs. Hence, one can safely use the anelastic approximation to simplify the calculations and reduce the numerical cost. Further, our results show that most of the inertial modes computed using the Boussinesq model are appreciably different from those calculated using the other models. The Boussinesq model reproduces identical inertial modes as the other models only for the \Rossby\ modes under uniform rotation, which are purely toroidal.
Otherwise, the absence of density stratification in the Boussinesq model leads to significant differences in non-toroidal modes, as it lacks the compressional $\beta$~effect. Therefore, a permissible setup for the studied solar inertial modes is the anelastic model with solar differential rotation because of the substantial effects of differential rotation on the eigenmodes.

There are still some issues remaining in the current modelling of the inertial modes. 
One prominent issue is the absence of the radiative interior below and the near-surface layer above the computational domain.
 {
Whether the anelastic formulation is valid for modelling the inertial modes in the radiative zone remains an open problem. The anelastic models have been found to fail to conserve energy for the gravity waves in the sub-adiabatic radiative interior, and pseudo-incompressible models have been shown to work better in such cases \citep{Brown2012ApJ, Vasil2013ApJ}. Thus, the anelastic approximation can likely affect the non-toroidal inertial modes in the radiative interior if they couple with gravity modes \citep[gravito-inertial modes, see, e.g.][]{Mathis2009A&A, Dintrans2000A&A}. Nonetheless, the anelastic formulation we use in our paper is expected to be valid for studying quasi-toroidal inertial modes in the radiative interior, as done through nonlinear numerical simulations \citep{Blume2024ApJ}.
}
On the other hand, it is well known that the anelastic approximation breaks down when modelling the solar convection in the near-surface layer where the stratification is strongly superadiabatic and the convective speed becomes as high as the sound speed \citep[e.g.][]{Nordlund2009LRSP}.
The extent to which the anelastic formulation can be used to model the low-frequency inertial modes in this near-surface layer requires future verification.

\begin{acknowledgements}
Author contributions: 
LG and XZ initiated this project, SM implemented all the equations in Dedalus and performed the computations, YB provided close supervision to validate the results, all authors discussed the results, SM wrote the initial draft, and all authors contributed to the final manuscript. 
We thank  R.~Cameron, P.~Dey, V.~Kannan, and J.~Schou for  helpful discussions. We also acknowledge the hospitality of Nordita, Stockholm, during the 2024 program on stellar convection. SM is a member of the International Max Planck Research School for Solar System Science at the University of G{\"o}ttingen. YB and LG acknowledge support from ERC Synergy Grant WHOLE SUN 810218. XZ acknowledges the financial support from the German Research Foundation (DFG) through grants 521319293, 540422505, and 550262949. The codes and data used in the manuscript are available in the Edmond database at \href{https://doi.org/10.17617/3.NP26AI}{https://doi.org/10.17617/3.NP26AI}.
\end{acknowledgements}

\bibliographystyle{aa}


\begin{appendix}

    \section{Sets of equations for the different models}\label{appendix:equations}
    Since there are subtle differences in the equations for the different models, we explicitly state the equations for the different models. The terms in the equations below are explained in \S~\ref{Model_descriptions}. 
    
    \subsection{Compressible model}
    The complete set of linearized equations for the compressible model is:
    \begin{multline}\label{eq:a1}
            \frac{\partial \textbf{u}}{\partial t}+\textbf{v}_0\cdot\nabla \textbf{u}+\textbf{u}\cdot\nabla \textbf{v}_0+2 \Omega_0\,\textbf{e}_z\times \textbf{u} +\mathbf{\nabla}\left(\frac{p_1}{\rho_0}\right) \\
     -\frac{s_1}{c_p}g\,\textbf{e}_r  - \frac{1}{ \rho_0}  \nabla\cdot \left( {\rho_0{\nu \left(\nabla (\textbf{u}) + \nabla (\textbf{u}^T) - \frac{2}{3}(\nabla \cdot \textbf{u}) \overleftrightarrow{I}\right)}}\right)=0,
    \end{multline}
    \begin{equation}\label{eq:a2}
    \frac{\partial s_1}{\partial t} + \textbf{v}_0\cdot\nabla s_1 + u_r\frac{\partial s_0}{\partial r} + \frac{u_{\theta}}{r} \frac{\partial s_0}{\partial \theta} - \frac{1}{ \rho_0 T_0}\nabla \cdot ( \kappa \rho_0 T_0 \nabla s_1)=0,
\end{equation}
\begin{equation}\label{eq:a3}
    \frac{\partial \rho_1}{\partial t}+\mathbf{\nabla}\cdot(\rho_0 \textbf{u})+\mathbf{\nabla}\cdot(\rho_1 \textbf{v}_0)=0,
\end{equation}
\begin{equation}\label{eq:a4}
    \frac{p_1}{p_0}=\gamma\frac{\rho_1}{\rho_0}+\frac{s_1}{c_v}.
\end{equation}
 For the compressible model, we solve Eqs.~\eqref{eq:a1} -- \eqref{eq:a3} with the constraint Eq.~\eqref{eq:a4}.

\subsection{Anelastic approximation}
 {We use the anelastic formulation, which conserves energy in both the convection and radiative zones and is recommended by \cite{Brown2012ApJ}.} The linearized equations for the anelastic model are:
 \begin{multline}\label{eq:a5}
            \frac{\partial \textbf{u}}{\partial t}+\textbf{v}_0\cdot\nabla \textbf{u}+\textbf{u}\cdot\nabla \textbf{v}_0+2 \Omega_0\,\textbf{e}_z\times \textbf{u} +\mathbf{\nabla}\left(\frac{p_1}{\rho_0}\right) \\
     -\frac{s_1}{c_p}g\,\textbf{e}_r  - \frac{1}{ \rho_0}  \nabla\cdot \left( {\rho_0{\nu \left(\nabla (\textbf{u}) + \nabla (\textbf{u}^T) - \frac{2}{3}(\nabla \cdot \textbf{u}) \overleftrightarrow{I}\right)}}\right)=0,
    \end{multline}
\begin{equation}\label{eq:a6}
    \frac{\partial s_1}{\partial t} + \textbf{v}_0\cdot\nabla s_1 + u_r\frac{\partial s_0}{\partial r} + \frac{u_{\theta}}{r} \frac{\partial s_0}{\partial \theta} - \frac{1}{ \rho_0 T_0}\nabla \cdot ( \kappa \rho_0 T_0 \nabla s_1)=0,
\end{equation}
\begin{equation}\label{eq:a7}
   \mathbf{\nabla}\cdot(\rho_0 \textbf{u})=0.
\end{equation}
In this case, we solve Eqs.~\eqref{eq:a5} -- \eqref{eq:a6}, with the anelastic constraint Eq.~\eqref{eq:a7}. In the Navier-Stokes equation, we implicitly assume that it follows the equation of state \eqref{eq:a4}, but we do not use it as an additional constraint when solving the eigenvalue problem. 

\subsection{Boussinesq approximation}
For the Boussinesq model, the linearized model equations are:
\begin{multline}\label{eq:a8}
            \frac{\partial \textbf{u}}{\partial t}+\textbf{v}_0\cdot\nabla \textbf{u}+\textbf{u}\cdot\nabla \textbf{v}_0+2 \Omega_0\,\textbf{e}_z\times \textbf{u} +\frac{1}{\rho_m}\mathbf{\nabla}{p_1}  \\
     -\frac{s_1}{c_p}g\,\textbf{e}_r  -   \nabla\cdot \left( {{\nu \left(\nabla (\textbf{u}) + \nabla (\textbf{u}^T) \right)}}\right)=0,
    \end{multline}
\begin{equation}\label{eq:a9}
    \frac{\partial s_1}{\partial t} + \textbf{v}_0\cdot\nabla s_1  + \frac{u_{\theta}}{r} \frac{\partial s_0}{\partial \theta} - \frac{1}{T_0}\nabla \cdot ( \kappa T_0   \nabla s_1)=0 ,
\end{equation}
\begin{equation}
\label{eq:a10}
    \nabla\cdot \textbf{u}=0 .
\end{equation}
Here, we again solve Eqs.~\eqref{eq:a8} -- \eqref{eq:a9} with the incompressibility constraint Eq.~\eqref{eq:a10}. Here,  $\rho_m$ is the mean density of the solar convection zone from the Solar model S (see Fig.~\ref{fig:1}). Unlike \cite{Triana2022ApJL}, we include the Boussinesq approximation, which considers the effect of density perturbations only in the buoyancy force term. The equations for the Boussinesq approximation are derived for a compressible fluid following procedures similar to those in \cite{Spiegel1960ApJ}.

We use the same background temperature $T_0$, which is almost adiabatic in the convection zone and taken from model S, for all three models. We consider no deviation from the adiabatic temperature gradient for simplicity and to match the conditions across all models. We also use the same latitudinal entropy gradient from thermal wind balance in all three models, given by
\begin{equation}
    \frac{\partial s_0}{\partial\theta} = r^2 \sin{\theta} \frac{g}{c_p}  \frac{\partial (\Omega^2)}{\partial z}.
\end{equation}

\section{Analysis of linearized vorticity equation for the  inertial modes}\label{appendix:Vorticity}
\begin{figure*}
    \centering
    \includegraphics[width=0.95\textwidth]{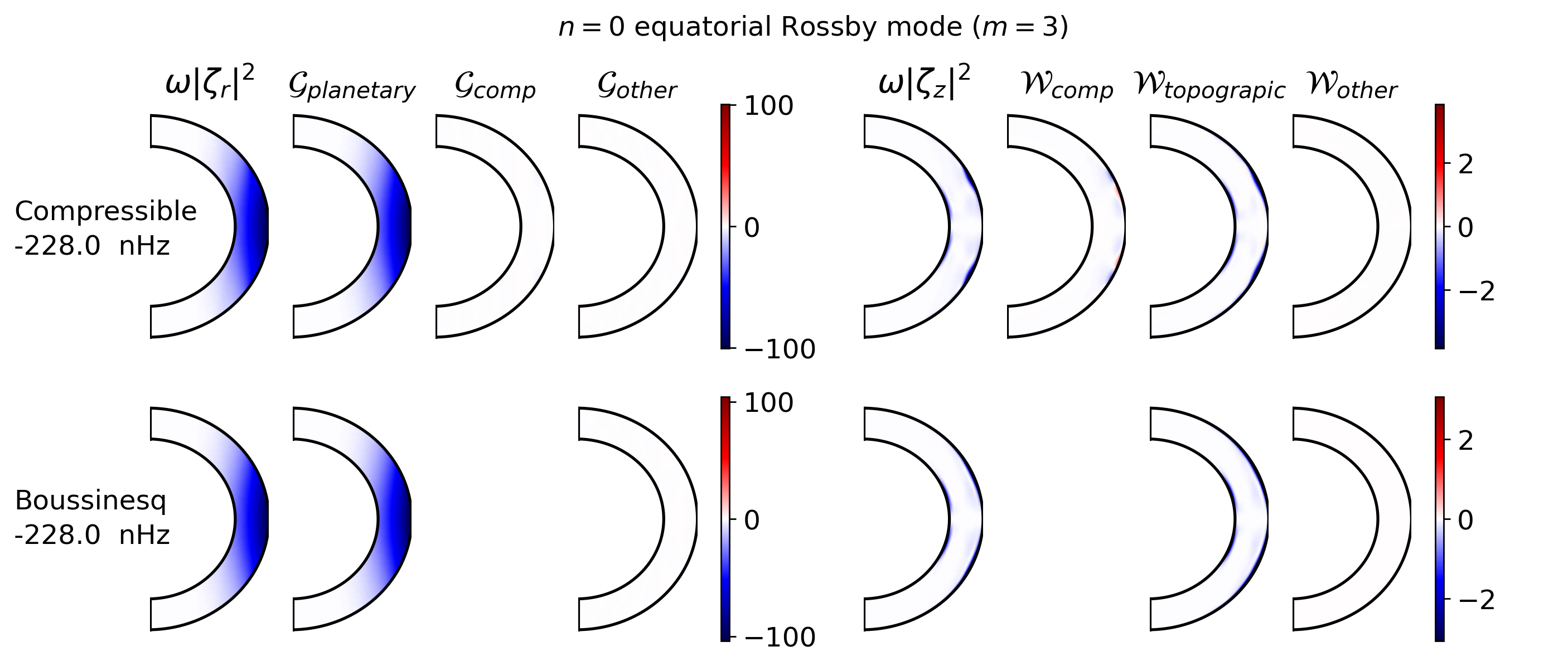}
    \caption{Relative importance of planetary $\beta$~effect ($\Re[ \mathcal{G}_{ \rm planetary}]$), compressional $\beta$~effect ($\Re[\mathcal{G}_{ \rm comp}], \Re[\mathcal{W}_{ \rm comp}]$), topographical $\beta$~effect ($\Re[\mathcal{W}_{ \rm topographic}]$) and other terms in the vorticity equation to determine the propagation and frequency of the \Rossby\ mode ($m=3$) in the compressible and Boussinesq models, under uniform rotation. Refer to Eqs.~\eqref{beta-G} - \eqref{beta-W} for the definition of the various quantities. A negative quantity implies that the associated physical effect promotes retrograde propagation, while a positive quantity implies that its physical effect promotes prograde propagation. The respective frequencies of the modes are specified on the left.  We normalize the eigenfunctions from the different models to have the same integrated kinetic energy density such that the maximum absolute value of $\Re[\omega|\zeta_r|^2]$ is $100$ for the compressible model.}
    \label{fig:8}
\end{figure*}

\begin{figure*}
    \centering
    \includegraphics[width=0.95\textwidth]{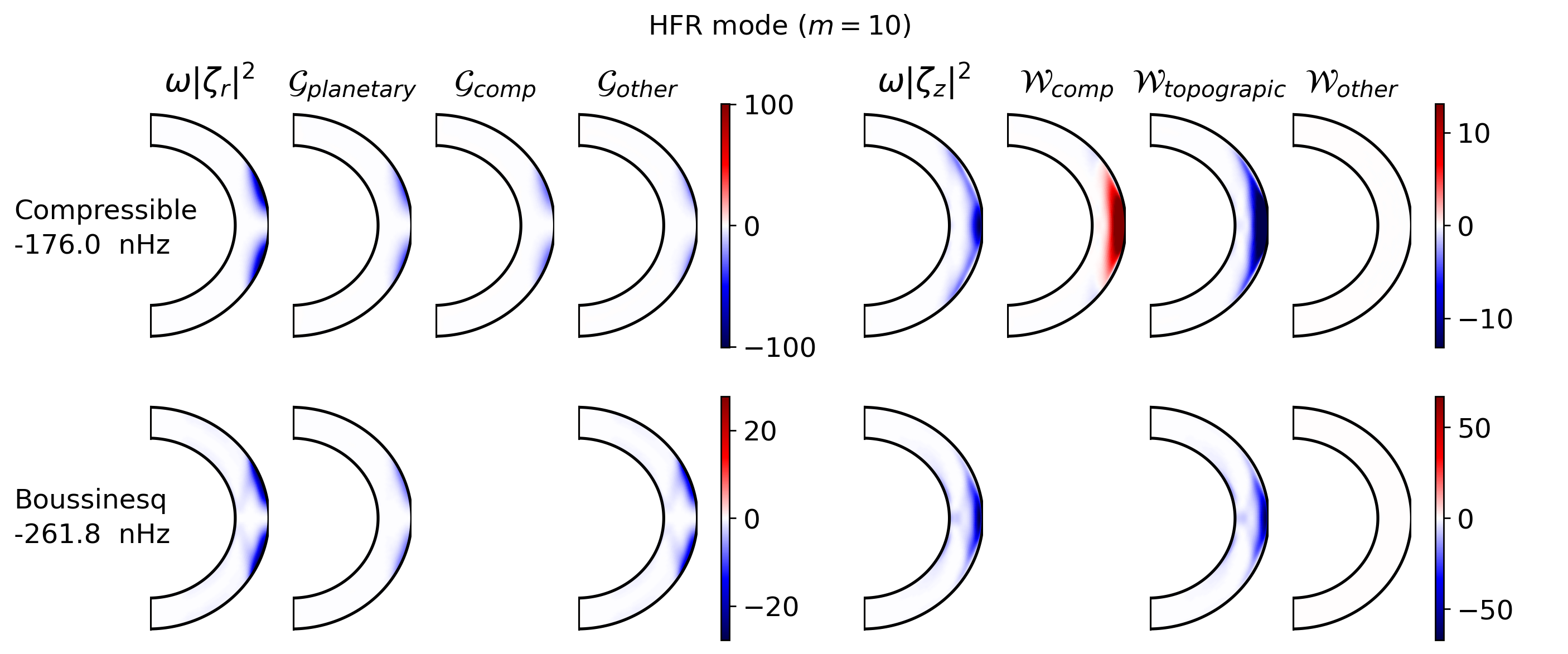}
    \caption{Relative importance of planetary $\beta$~effect ($\Re[ \mathcal{G}_{ \rm planetary}]$), compressional $\beta$~effect ($\Re[\mathcal{G}_{ \rm comp}], \Re[\mathcal{W}_{ \rm comp}]$), topographical $\beta$~effect ($\Re[\mathcal{W}_{ \rm topographic}]$) and other terms in the vorticity equation to determine the propagation and frequency of the \HFR\ mode ($m=10$) in the compressible and Boussinesq models, under uniform rotation. We use the same definition and interpretation for the different quantities as in Fig.~\ref{fig:8}. The respective frequencies of the modes are specified on the left. The eigenfunctions have the same normalization as in Fig.~\ref{fig:8}.}
    \label{fig:9}
\end{figure*}

\begin{figure*}
    \centering
    \includegraphics[width=0.95\textwidth]{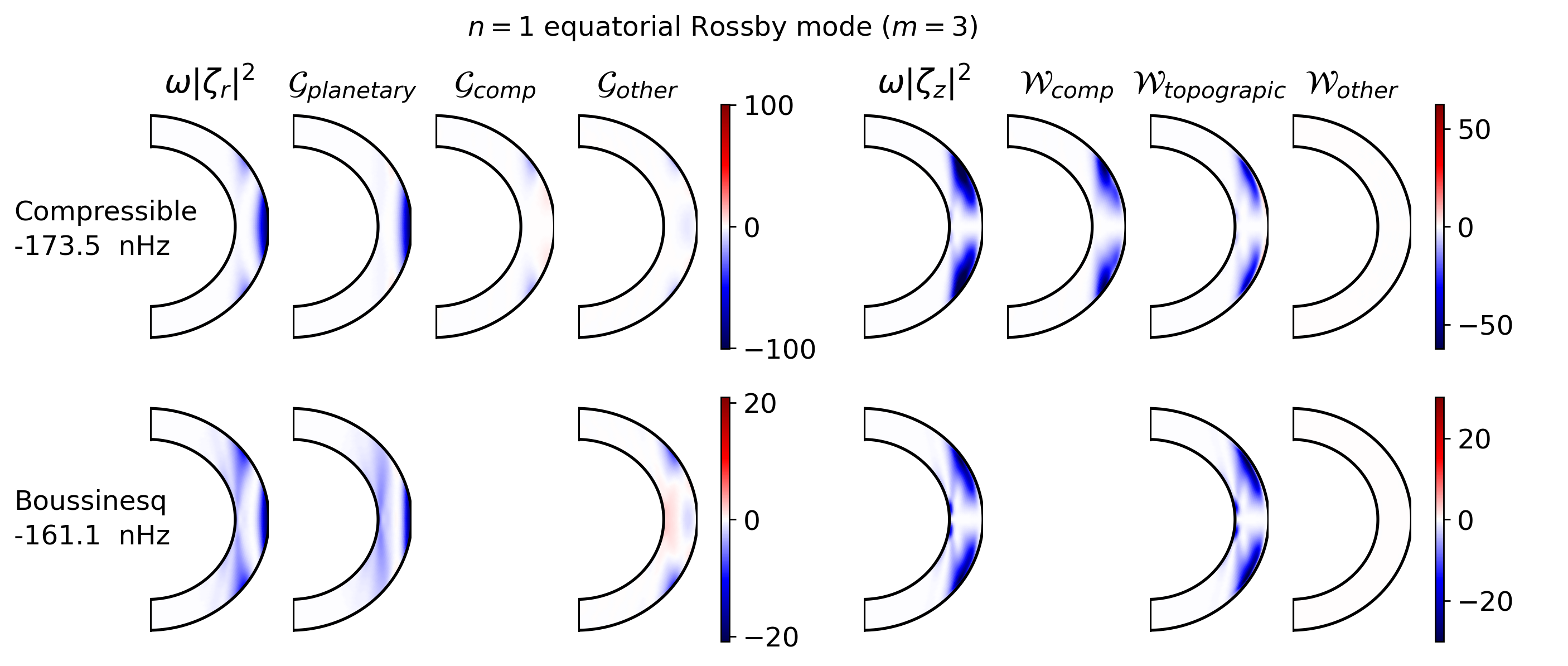}
    \caption{Relative importance of planetary $\beta$~effect ($\Re[ \mathcal{G}_{ \rm planetary}]$), compressional $\beta$~effect ($\Re[\mathcal{G}_{ \rm comp}], \Re[\mathcal{W}_{ \rm comp}]$), topographical $\beta$~effect ($\Re[\mathcal{W}_{ \rm topographic}]$) and other terms in the vorticity equation to determine the propagation and frequency of the \Rossbyno\ mode ($m=3$) in the compressible and Boussinesq models, under uniform rotation.  We use the same definition and interpretation for the different quantities as in Fig.~\ref{fig:8}. The respective frequencies of the modes are specified on the left. The eigenfunctions have the same normalization as in Fig.~\ref{fig:8}. }
    \label{fig:10}
\end{figure*}

\begin{figure*}
    \centering
    \includegraphics[width=0.95\textwidth]{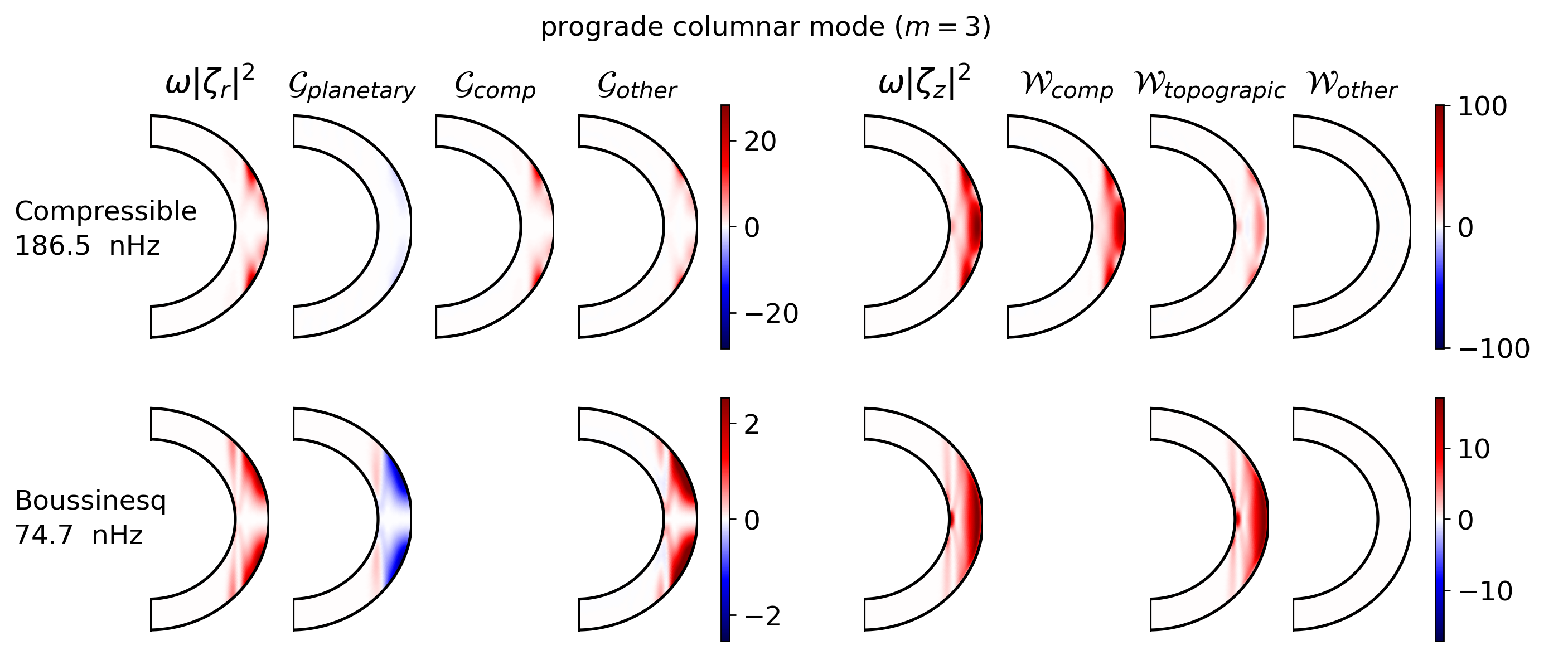}
    \caption{Relative importance of planetary $\beta$~effect ($\Re[ \mathcal{G}_{ \rm planetary}]$), compressional $\beta$~effect ($\Re[\mathcal{G}_{ \rm comp}], \Re[\mathcal{W}_{ \rm comp}]$), topographical $\beta$~effect ($\Re[\mathcal{W}_{ \rm topographic}]$) and other terms in the vorticity equation to determine the propagation and frequency of the prograde columnar mode ($m=3$) in the compressible and Boussinesq models, under uniform rotation. We use the same definition and interpretation for the different quantities as in Fig.~\ref{fig:8}. The respective frequencies of the modes are specified on the left. We normalize the eigenfunctions from the different models to have the same integrated kinetic energy density such that the maximum of $\Re[\omega|\zeta_z|^2]$ is 100 for the compressible model.}
    \label{fig:11}
\end{figure*}

In this appendix, we present detailed analyses of the terms in the radial and $z$ vorticity equations (Eqs.~(\ref{eq:r-vorticity-final}) and (\ref{eq:z-vorticity-final})).
Figure~\ref{fig:8} compares the significance of the left and right-hand side terms of the linearized vorticity equations for the \Rossby\ modes with $m=3$ from the compressible and Boussinesq models under uniform rotation.
The planetary $\beta$~effect represented by $\Re[\mathcal{G}_{\rm planetary}]$ drives the retrograde propagation of these modes.
Since the \Rossby\ modes are toroidal under uniform rotation, no difference is observed between the compressible and Boussinesq results.
Figure~\ref{fig:9} shows the results for the $m=10$ \HFR\ mode.
It indicates that the retrograde propagation of the \HFR\ modes is driven by combined effects of $\mathcal{G}_{\rm planetary}$, $\mathcal{G}_{\rm comp}$, $\mathcal{G}_{\rm other}$, and $\mathcal{W}_{\rm topographic}$.
The radial motions associated with the $z$-vorticity induce the compressional $\beta$~effect $\Re [  \mathcal{W}_{\rm comp}]$, which counteracts the other terms and promotes prograde propagation.
In the Boussinesq model, the mode frequencies become more retrograde due to the absence of $\mathcal{W}_{\rm comp}$.
Figure~\ref{fig:10} presents the results for the \Rossbyno\ mode with $m=3$.
In contrast to the \Rossby\ mode, the \Rossbyno\ mode exhibits both radial and $z$ vortical motions.
For the radial component of the vorticity $\zeta_{r}$, the planetary $\beta$~effect is the primary driver of their retrograde propagation.
We also show that the topographic $\beta$~effect represented by $\Re[\mathcal{W}_{\rm topographic}]$ enhances the retrograde propagation of their $z$-vorticity $\zeta_{z}$. 
In the Boussinesq model, the mode frequencies become slightly more prograde due to the absence of the retrograde-promoting $\mathcal{W}_{\rm comp}$.
Figure~\ref{fig:11} shows the results for the $m=3$ prograde columnar mode.
The $z$-vortical motions dominate over the radial-vortical motions in these modes.
In the compressible model, their prograde propagation is primarily driven by the compressional $\beta$~effect represented by $\Re[\mathcal{W}_{\rm comp}]$ and weakly influenced by the topographic $\beta$~effect represented by $\Re[\mathcal{W}_{\rm topographic}]$.
In the Boussinesq model, $\Re[\mathcal{W}_{\rm topographic}]$ is the sole driver of their prograde propagation.
Furthermore, the lack of diverging (converging) tendency of upflows (downflows) in the Boussinesq model amplifies the radial vortical motions at the surface, thereby enhancing the planetary $\beta$~effect $\Re[\mathcal{G}_{\rm planetary}]$, which reinforces the retrograde mode propagation.
Thus, in the Boussinesq model, the prograde columnar modes exhibit significantly lower frequencies, as shown in Fig.~\ref{fig:3}.

\FloatBarrier

\section{Quantitative comparison of anelastic and Boussinesq models with respect to compressible model}\label{appendix:quantitative}

\begin{table}
\small
    \centering
\caption{Frequencies and non-toroidicity of the \Rossby\ modes under uniform rotation.}
    \begin{tabular}{c c c c c}
\hline\hline
\multirow{2}{*}{$m$} & Compressible  & Compr. - &  Compr. - & \multirow{2}{*}{$\Gamma$}\\
&$\omega/2\pi$ (nHz)&Anelastic &Boussinesq & \\
\hline
1 & $-455.9944$ & $0.0056$ & $0.0056$ & $0.0000$ \\
2 & $-303.9977$ & $0.0018$ & $0.0020$ & $0.0009$ \\
3 & $-227.9918$ & $0.0008$ & $0.0023$ & $0.0019$ \\
4 & $-182.3657$ & $0.0004$ & $0.0061$ & $0.0031$ \\
5 & $-151.9042$ & $0.0002$ & $0.0175$ & $0.0045$ \\
6 & $-130.0859$ & $0.0001$ & $0.0388$ & $0.0060$ \\
7 & $-113.6626$ & $0.0001$ & $0.0661$ & $0.0076$ \\
8 & $-100.8461$ & $0.0001$ & $0.0907$ & $0.0090$ \\
9 & $-90.5729$ & $0.0001$ & $0.1043$ & $0.0103$ \\
10 & $-82.1683$ & $0.0000$ & $0.1026$ & $0.0113$ \\
11 & $-75.1793$ & $0.0000$ & $0.0855$ & $0.0120$ \\
12 & $-69.2873$ & $0.0000$ & $0.0562$ & $0.0125$ \\
13 & $-64.2603$ & $0.0000$ & $0.0198$ & $0.0127$ \\
14 & $-59.9249$ & $0.0000$ & $-0.0192$ & $0.0128$ \\
15 & $-56.1494$ & $0.0000$ & $-0.0569$ & $0.0127$ \\
16 & $-52.8323$ & $0.0000$ & $-0.0911$ & $0.0125$ \\
\hline
\end{tabular}
    \tablefoot{The first column denotes the azimuthal order $m$. The second column reports the frequency ($\omega$) of the modes in the Carrington frame as computed using the compressible model. The third and fourth columns report the differences in the frequencies (in nHz) for the anelastic and the Boussinesq models, respectively, relative to the compressible model. The fifth column denotes the non-toroidicity $\Gamma$ as computed for the modes from the compressible model using Eq.~\eqref{Non-toroidicity}.}
    \label{tab:1}
\end{table}

\begin{table}
\small
    \centering
    \caption{Frequencies and non-toroidicity of the \HFR\ modes under uniform rotation, with the same notation as in Table~\ref{tab:1}.}
    \begin{tabular}{c c c c c}
\hline\hline
\multirow{2}{*}{$m$} & Compressible  & Compr. - &  Compr. - & \multirow{2}{*}{$\Gamma$}\\
&$\omega/2\pi$ (nHz)&Anelastic &Boussinesq & \\
\hline
2 & $-362.0800$ & $0.0009$ & $70.0505$ & $0.2352$ \\
3 & $-329.3193$ & $0.0007$ & $69.9538$ & $0.2090$ \\
4 & $-298.4230$ & $0.0005$ & $73.6560$ & $0.1999$ \\
5 & $-270.5835$ & $0.0004$ & $77.7805$ & $0.2011$ \\
6 & $-246.0950$ & $0.0003$ & $81.0767$ & $0.2058$ \\
7 & $-224.7627$ & $0.0002$ & $83.4430$ & $0.2110$ \\
8 & $-206.2202$ & $0.0002$ & $84.9281$ & $0.2159$ \\
9 & $-190.0747$ & $0.0002$ & $85.6682$ & $0.2203$ \\
10 & $-175.9641$ & $0.0001$ & $85.8339$ & $0.2242$ \\
11 & $-163.5766$ & $0.0001$ & $85.5806$ & $0.2279$ \\
12 & $-152.6507$ & $0.0001$ & $85.0296$ & $0.2315$ \\
13 & $-142.9701$ & $0.0001$ & $84.2688$ & $0.2349$ \\
14 & $-134.3562$ & $0.0001$ & $83.3587$ & $0.2381$ \\
15 & $-126.6609$ & $0.0001$ & $82.3406$ & $0.2412$ \\
16 & $-119.7606$ & $0.0000$ & $81.2419$ & $0.2440$ \\
\hline
\end{tabular}
    \label{tab:2}
\end{table}

\begin{table}
\small
    \centering
    \caption{Frequencies and non-toroidicity of the \Rossbyno\ modes under uniform rotation, with the same notation as in Table~\ref{tab:1}.}
    \begin{tabular}{c c c c c}
\hline\hline
\multirow{2}{*}{$m$} & Compressible  & Compr. - &  Compr. - & \multirow{2}{*}{$\Gamma$}\\
&$\omega/2\pi$ (nHz)&Anelastic &Boussinesq & \\
\hline
0 & $-274.9041$ & $0.0011$ & $-28.7642$ & $0.1454$ \\
1 & $-238.8316$ & $0.0008$ & $-18.4009$ & $0.1102$ \\
2 & $-203.4461$ & $0.0005$ & $-14.6810$ & $0.0791$ \\
3 & $-173.4967$ & $0.0003$ & $-12.3822$ & $0.0591$ \\
4 & $-149.3700$ & $0.0002$ & $-10.5521$ & $0.0468$ \\
5 & $-130.1043$ & $0.0001$ & $-9.0625$ & $0.0394$ \\
6 & $-114.6432$ & $0.0001$ & $-7.8716$ & $0.0350$ \\
7 & $-102.1167$ & $0.0001$ & $-6.9313$ & $0.0323$ \\
8 & $-91.8530$ & $0.0001$ & $-6.1867$ & $0.0307$ \\
9 & $-83.3416$ & $0.0000$ & $-5.5842$ & $0.0296$ \\
10 & $-76.1963$ & $0.0000$ & $-5.0785$ & $0.0288$ \\
11 & $-70.1272$ & $0.0000$ & $-4.6374$ & $0.0282$ \\
12 & $-64.9176$ & $0.0000$ & $-4.2419$ & $0.0277$ \\
13 & $-60.4051$ & $0.0000$ & $-3.8832$ & $0.0273$ \\
14 & $-56.4665$ & $0.0000$ & $-3.5580$ & $0.0270$ \\
15 & $-53.0063$ & $0.0000$ & $-3.2646$ & $0.0267$ \\
16 & $-49.9489$ & $0.0000$ & $-3.0020$ & $0.0263$ \\
\hline
\end{tabular}
    \label{tab:3}
\end{table}

\begin{table}
\small
    \centering
    \caption{Frequencies and non-toroidicity of the prograde columnar modes under uniform rotation, with the same notation as in Table~\ref{tab:1}.}
    \begin{tabular}{c c c c c}
\hline\hline
\multirow{2}{*}{$m$} & Compressible  & Compr. - &  Compr. - & \multirow{2}{*}{$\Gamma$}\\
&$\omega/2\pi$ (nHz)&Anelastic &Boussinesq & \\
\hline
0 & $0.0000$ & $-0.0000$ & $0.0000$ & $0.0011$ \\
1 & $67.7673$ & $-0.0004$ & $38.5542$ & $0.0656$ \\
2 & $130.7501$ & $-0.0006$ & $75.8710$ & $0.1259$ \\
3 & $186.5260$ & $-0.0008$ & $111.8153$ & $0.1806$ \\
4 & $235.2843$ & $-0.0009$ & $146.3101$ & $0.2272$ \\
5 & $276.9278$ & $-0.0009$ & $178.2647$ & $0.2669$ \\
6 & $310.8563$ & $-0.0009$ & $206.0269$ & $0.3029$ \\
7 & $338.7767$ & $-0.0009$ & $230.3947$ & $0.3361$ \\
8 & $362.0612$ & $-0.0009$ & $252.0103$ & $0.3712$ \\
9 & $380.8134$ & $-0.0008$ & $270.4158$ & $0.4101$ \\
10 & $393.3064$ & $-0.0008$ & $283.4660$ & $0.4456$ \\
11 & $402.3409$ & $-0.0007$ & $293.6587$ & $0.4843$ \\
12 & $407.2806$ & $-0.0006$ & $300.1433$ & $0.5109$ \\
13 & $410.8534$ & $-0.0006$ & $305.4989$ & $0.5360$ \\
14 & $412.0980$ & $-0.0005$ & $308.6629$ & $0.5557$ \\
15 & $412.1763$ & $-0.0005$ & $310.7292$ & $0.5672$ \\
16 & $411.6828$ & $-0.0005$ & $312.2481$ & $0.5744$ \\
\hline
\end{tabular}
    \label{tab:4}
\end{table}

\begin{table}
\small
    \centering
\caption{Frequencies and non-toroidicity of the \Rossby\ modes under solar differential rotation, with the same notation as in Table~\ref{tab:1}.}
    \begin{tabular}{c c c c c}
\hline\hline
\multirow{2}{*}{$m$} & Compressible  & Compr. - &  Compr. - & \multirow{2}{*}{$\Gamma$}\\
&$\omega/2\pi$ (nHz)&Anelastic &Boussinesq & \\
\hline
1 & $-455.9953$ & $0.0044$ & $0.0047$ & $0.0011$ \\
2 & $-331.8168$ & $-0.0020$ & $-2.5239$ & $0.0070$ \\
3 & $-262.0695$ & $-0.0134$ & $-4.1339$ & $0.0135$ \\
4 & $-230.0002$ & $0.0083$ & $-5.6185$ & $0.1134$ \\
5 & $-197.0644$ & $0.0017$ & $-6.3740$ & $0.1030$ \\
6 & $-171.2207$ & $0.0024$ & $7.8208$ & $0.0954$ \\
7 & $-130.8246$ & $-0.0040$ & $0.0611$ & $0.0882$ \\
8 & $-122.1143$ & $-0.0014$ & $1.5886$ & $0.0847$ \\
9 & $-114.7690$ & $-0.0016$ & $5.4175$ & $0.0893$ \\
10 & $-109.5107$ & $-0.0021$ & $9.2889$ & $0.0984$ \\
11 & $-105.7246$ & $-0.0025$ & $12.9504$ & $0.1087$ \\
12 & $-102.9795$ & $-0.0028$ & $16.4166$ & $0.1190$ \\
13 & $-100.9864$ & $-0.0031$ & $19.6566$ & $0.1291$ \\
14 & $-99.5498$ & $-0.0032$ & ... & $0.1384$ \\
15 & $-98.5366$ & $-0.0034$ & ... & $0.1459$ \\
16 & $-97.8500$ & $-0.0035$ & ... & $0.1510$ \\
\hline
\end{tabular}
    \label{tab:5}
\end{table}

\begin{table}
\small
    \centering
    \caption{Frequencies and non-toroidicity of the high-latitude modes under solar differential rotation, with the same notation as in Table~\ref{tab:1}.}
    \begin{tabular}{c c c c c}
\hline\hline
\multirow{2}{*}{$m$} & Compressible  & Compr. - &  Compr. - & \multirow{2}{*}{$\Gamma$}\\
&$\omega/2\pi$ (nHz)&Anelastic &Boussinesq & \\
\hline
0 & $-0.0000$ & $-0.0000$ & $-0.0000$ & $0.0011$ \\
1 & $-95.6996$ & $0.0034$ & $12.2996$ & $0.0250$ \\
2 & $-184.3896$ & $0.0058$ & $29.1408$ & $0.0641$ \\
3 & $-270.5509$ & $0.0200$ & $45.0518$ & $0.1163$ \\
4 & $-354.2757$ & $0.0074$ & $57.2498$ & $0.1713$ \\
5 & $-434.7199$ & $0.0068$ & $64.7203$ & $0.2256$ \\
6 & $-511.4853$ & $0.0050$ & $68.0383$ & $0.2762$ \\
7 & $-584.4776$ & $0.0029$ & $68.4952$ & $0.3214$ \\
8 & $-653.8597$ & $0.0007$ & $67.0029$ & $0.3609$ \\
\hline
\end{tabular}
    \label{tab:6}
\end{table}

\begin{table}
\small
    \centering
    \caption{Frequencies and non-toroidicity of the \HFR\ modes under solar differential rotation, with the same notation as in Table~\ref{tab:1}.}
    \begin{tabular}{c c c c c}
\hline\hline
\multirow{2}{*}{$m$} & Compressible  & Compr. - &  Compr. - & \multirow{2}{*}{$\Gamma$}\\
&$\omega/2\pi$ (nHz)&Anelastic &Boussinesq & \\
\hline
2 & $-380.0809$ & $0.0018$ & $70.7218$ & $0.2002$ \\
3 & $-350.1367$ & $0.0027$ & $70.6547$ & $0.2020$ \\
4 & $-320.1066$ & $0.0039$ & $80.6738$ & $0.2125$ \\
5 & $-293.3019$ & $0.0053$ & $84.9598$ & $0.2234$ \\
6 & $-270.2756$ & $0.0068$ & $84.8862$ & $0.2428$ \\
7 & $-251.0942$ & $0.0083$ & $81.9565$ & $0.2587$ \\
8 & $-235.3597$ & $0.0096$ & $77.3598$ & $0.2753$ \\
9 & $-222.3900$ & $0.0104$ & $71.7780$ & $0.2862$ \\
10 & $-211.2815$ & $0.0097$ & $65.9498$ & $0.2830$ \\
11 & $-202.2048$ & $0.0075$ & $59.8404$ & $0.2738$ \\
12 & $-195.3937$ & $0.0055$ & $54.3140$ & $0.2686$ \\
13 & $-190.3007$ & $0.0042$ & $50.9713$ & $0.2682$ \\
14 & $-186.2602$ & $0.0035$ & $49.7995$ & $0.2708$ \\
15 & $-182.8271$ & $0.0030$ & $49.7990$ & $0.2747$ \\
16 & $-179.7445$ & $0.0025$ & $50.2232$ & $0.2790$ \\
\hline
\end{tabular}
    \label{tab:7}
\end{table}

\begin{table}
\small
    \centering
    \caption{Frequencies and non-toroidicity of the \Rossbyno\ modes under solar differential rotation, with the same notation as in Table~\ref{tab:1}.}
    \begin{tabular}{c c c c c}
\hline\hline
\multirow{2}{*}{$m$} & Compressible  & Compr. - &  Compr. - & \multirow{2}{*}{$\Gamma$}\\
&$\omega/2\pi$ (nHz)&Anelastic &Boussinesq & \\
\hline
0 & $-288.5330$ & $0.0004$ & $-31.1401$ & $0.1033$ \\
1 & $-246.9782$ & $0.0007$ & $-18.5591$ & $0.1079$ \\
2 & $-210.5457$ & $0.0007$ & $-8.0265$ & $0.0808$ \\
3 & $-182.0516$ & $-0.0001$ & $2.7757$ & $0.0908$ \\
4 & $-151.9285$ & $-0.0003$ & $5.5443$ & $0.0652$ \\
5 & $-129.2566$ & $-0.0007$ & $2.9265$ & $0.0592$ \\
6 & $-108.2980$ & $-0.0023$ & $-1.8894$ & $0.0531$ \\
7 & $-89.0665$ & $-0.0047$ & $4.5684$ & $0.0439$ \\
8 & $-73.4214$ & $-0.0080$ & $15.0778$ & $0.0378$ \\
9 & $-60.4377$ & $-0.0117$ & $25.3317$ & $0.0399$ \\
10 & $-49.7833$ & $-0.0136$ & $33.7863$ & $0.0503$ \\
11 & $-41.1950$ & $-0.0128$ & $40.1575$ & $0.0652$ \\
12 & $-34.2704$ & $-0.0104$ & $44.6014$ & $0.0834$ \\
13 & $-28.4929$ & $-0.0083$ & $47.4666$ & $0.1044$ \\
14 & $-23.3225$ & $-0.0071$ & $49.2240$ & $0.1260$ \\
15 & $-18.4022$ & $-0.0064$ & $50.2401$ & $0.1466$ \\
16 & $-13.5597$ & $-0.0059$ & $50.7363$ & $0.1661$ \\
\hline
\end{tabular}
    \label{tab:8}
\end{table}

\begin{table}
\small
    \centering
    \caption{Frequencies and non-toroidicity of the prograde columnar modes under solar differential rotation, with the same notation as in Table~\ref{tab:1}.}
    \begin{tabular}{c c c c c}
\hline\hline
\multirow{2}{*}{$m$} & Compressible  & Compr. - &  Compr. - & \multirow{2}{*}{$\Gamma$}\\
&$\omega/2\pi$ (nHz)&Anelastic &Boussinesq & \\
\hline
0 & $-0.0000$ & $-0.0000$ & $-0.0000$ & $0.0058$ \\
1 & $76.7750$ & $0.0002$ & $49.5405$ & $0.0650$ \\
2 & $147.6312$ & $0.0004$ & $98.1219$ & $0.1252$ \\
3 & $210.8393$ & $0.0009$ & $145.6926$ & $0.1824$ \\
4 & $267.4129$ & $0.0017$ & $192.3226$ & $0.2262$ \\
5 & $316.0582$ & $0.0025$ & $232.8312$ & $0.2566$ \\
6 & $370.4610$ & $0.0029$ & $279.3092$ & $0.3085$ \\
7 & $400.2431$ & $0.0029$ & $299.3421$ & $0.3563$ \\
8 & $423.8501$ & $0.0033$ & $315.1826$ & $0.3817$ \\
9 & $443.9587$ & $0.0034$ & $327.5663$ & $0.3951$ \\
10 & $461.9409$ & $0.0033$ & $337.8791$ & $0.4052$ \\
11 & $478.4130$ & $0.0031$ & $346.8144$ & $0.4158$ \\
12 & $493.5135$ & $0.0028$ & $354.5033$ & $0.4273$ \\
13 & $507.3183$ & $0.0026$ & $361.0170$ & $0.4387$ \\
14 & $519.9892$ & $0.0025$ & ... & $0.4496$ \\
15 & $531.7273$ & $0.0024$ & ... & $0.4598$ \\
16 & $542.7219$ & $0.0024$ & ... & $0.4692$ \\
\hline
\end{tabular}
    \label{tab:9}
\end{table}

\begin{figure*}
    \centering
    \includegraphics[width=0.93\textwidth]{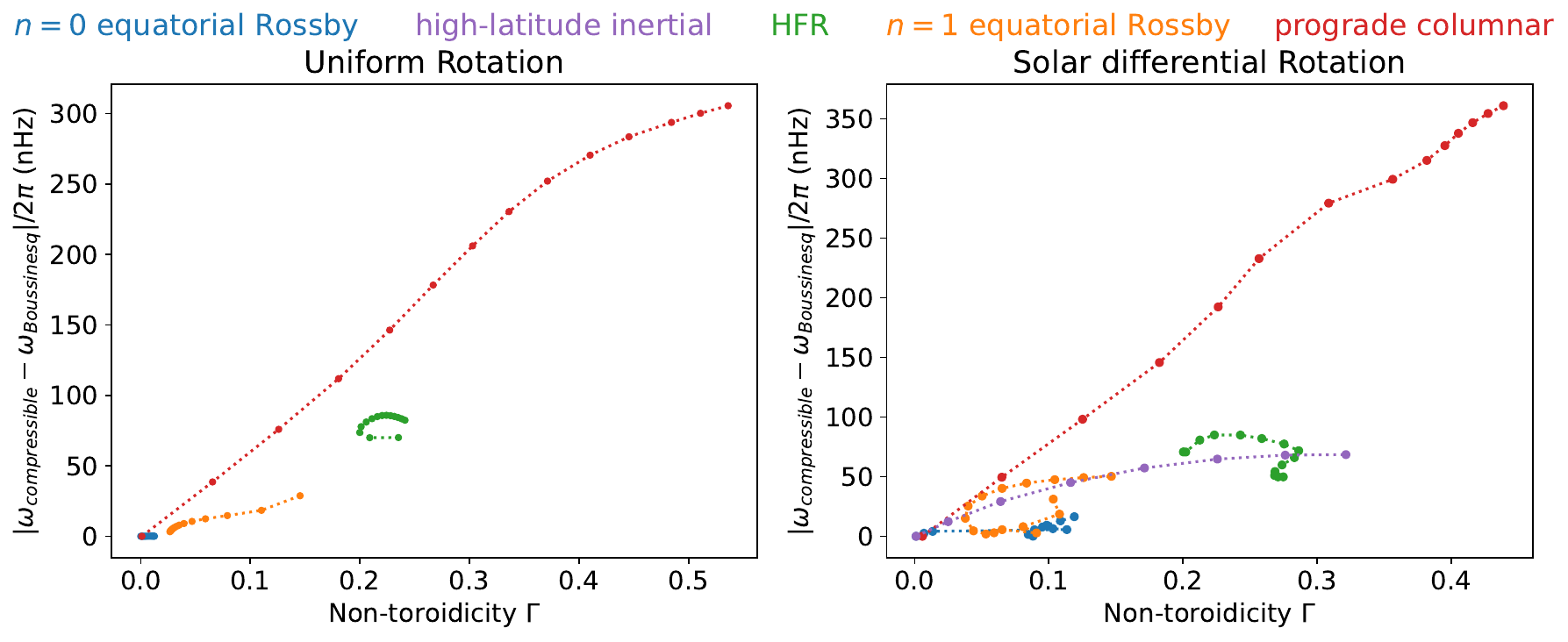}
    \caption{Effects of the non-toroidicity of the eigenmodes on the absolute difference between the real part of the eigenfrequencies for the compressible and Boussinesq models. The non-toroidicity $\Gamma$ of the eigenmodes is calculated using Eq.~\eqref{Non-toroidicity} for the compressible model. The left column corresponds to uniform rotation, while the right corresponds to solar differential rotation. The different colours denote the different modes, similar to Fig.~\ref{fig:5}. Points of the same colour correspond to modes with different azimuthal orders $m$.}
    \label{fig:12}
\end{figure*}

 The following Tables~\ref{tab:1} --  \ref{tab:4} present the frequencies $\omega$   for the three models and non-toroidicity $\Gamma$ of different modes under uniform rotation. Tables~\ref{tab:5} -- \ref{tab:9} report the same for the modes under solar differential rotation and its corresponding entropy gradient. The non-toroidicity, represented as $\Gamma$, is calculated using the eigenfunction for the compressible model with Eq.~\eqref{Non-toroidicity}. The tables indicate that the compressible and anelastic models yield the same frequencies accurate to $0.1$ nHz. In contrast, the Boussinesq model produces significantly different frequencies for almost all the modes except the \Rossby\ modes under uniform rotation. 
 The computation of the eigenmodes by the compressible model requires about 1.33 times the computational time needed for the anelastic model. Thus, the anelastic model reduces computational cost without altering the solar inertial modes.

Figure~\ref{fig:12} shows the non-toroidicity of the modes computed using the compressible model plotted against the absolute difference in frequencies between the Boussinesq and compressible models under uniform and solar differential rotation. In general, the frequency difference increases with the non-toroidicity of the different inertial modes. The trend is most pronounced for the prograde columnar modes under both uniform and differential rotation and the high-latitude modes under differential rotation. It is also evident for the \Rossbyno\ modes under uniform rotation. However, the frequency difference decreases with increasing azimuthal order for these modes (see Fig.~\ref{fig:3}). Under differential rotation, the non-toroidal effects of critical latitudes become more significant with increasing azimuthal order. These competing effects complicate the relationship between non-toroidicity and frequency difference for the \Rossbyno\ modes under differential rotation. The variations in frequency difference and non-toroidicity remain relatively small for the \HFR\ modes under uniform rotation. However, under differential rotation, critical latitudes introduce variations in the \HFR\ modes. The \Rossby\ modes are nearly toroidal and exhibit no frequency difference under uniform rotation. Under differential rotation, they acquire a slight non-toroidal character and exhibit small frequency differences due to the radial motions near critical latitudes at higher values of $m$. Although non-toroidicity provides a rough estimate of the differences in eigenmodes caused by the Boussinesq approximation, it is not a precise measure, as additional physical effects contribute to the characteristics of the inertial modes.

\FloatBarrier

\end{appendix}

\end{document}